\documentclass[aps,prd,eqsecnum,preprintnumbers,showpacs]{revtex4}
\usepackage{bm}
\usepackage{graphicx}
\begin{document}

\vspace*{-10.0mm} 
\thispagestyle{empty}

\rightline{\large\baselineskip20pt\rm\vbox to20pt{
\baselineskip14pt
 \hbox{YITP-03-81}}}

\vskip8mm
\begin{center}
{\large\bf
 Local conservation law and dark radiation in cosmological
braneworld}
\end{center}

\begin{center}
\large
Masato Minamitsuji$^{1,2}$
and 
Misao Sasaki$^2$
\end{center}

\begin{center}
{\em
$^1$Department of Earth and Space Science, Graduate School of Science,
\\
Osaka University, Toyonaka 560-0043, Japan}
\end{center}

\begin{center}
{\em
$^2$Yukawa Institute for Theoretical Physics,
\\
 Kyoto University, Kyoto 606-8502, Japan}
\end{center}

\begin{abstract}
In the context of the Randall-Sundrum (RS) single-brane
scenario, we discuss the bulk geometry and dynamics
of a cosmological brane in terms of 
the local energy conservation law which exists for 
the bulk that allows slicing with
a maximally symmetric 3-space. This
conservation law enables
us to define a local mass in the bulk.
We show that there is a unique generalization of the
dark radiation on the brane, which is given by
the local mass.
We find there also exists a conserved current associated
with the Weyl tensor, and the corresponding local charge,
which we call the Weyl charge,
is given by the sum of the local mass and a certain
linear combination of the components of the bulk
energy-momentum tensor. This expression of the Weyl charge
relates the local mass with the projected Weyl tensor, 
$E_{\mu\nu}$, which plays a central role in the geometrical
formalism of the RS braneworld.
On the brane, in particular, this gives a decomposition of
the projected Weyl tensor into the local mass
and the bulk energy-momentum tensor.
Then, as an application of these results,
we consider a null dust model for the bulk energy-momentum
tensor and discuss the black hole formation in the bulk.
We investigate the causal structure by identifying
the locus of the apparent horizon
and clarify possible brane trajectories in the bulk.
We find that the brane stays always outside the 
black hole as long as it is expanding.
We also find an upper bound on the value of the Hubble parameter
in terms of the matter energy density on the brane,
irrespective of the energy flux emitted
from the brane.
\end{abstract}


\date{\today}

\maketitle

\makeatletter
\makeatother

\section{Introduction}

The braneworld scenario has attracted much attention
in recent years \cite{Horava:1995qa}.
In this scenario, our universe is assumed to be on a
(mem)brane embedded in a higher dimensional spacetime.
There are many models of the braneworld scenario
and corresponding cosmologies.
One of them that has been extensively
studied is the braneworld cosmology based on a
model proposed by Randall and Sundrum \cite{Randall:1999vf},
in which a single positive tension brane exists in
a 5-dimensional spacetime (called the bulk) with negative
cosmological constant, the so-called RS2 model.
In this paper, we focus our discussion
on this single-brane model.

In many cases, the 5-dimensional bulk geometry
is assumed to be Anti-de Sitter (AdS) or AdS-Schwarzschild
\cite{Ida:1999ui, Binetruy:1999ut, Maartens:2003tw}:
\begin{eqnarray}
ds^2=-\Bigl(K+{r^2\over \ell^2}-{M_{0}\over r^2}\Bigr)dt^2 +
\Bigl(K+{r^2\over \ell^2}-{M_{0}\over r^2}\Bigr)^{-1}dr^2
+r^2 d\Omega_{(K,3)}^2,
\label{adssch}
\end{eqnarray}
where $\ell:=\sqrt{-6/\Lambda_{5}}$ is the AdS curvature radius,
$M_{0}$ is the black hole mass, and $d\Omega_{(K,3)}^2$ is
the maximally symmetric (constant curvature)
3-space with $K=-1$, $0$ or $+1$. 
The brane trajectory in the bulk, $(t,r)=\bigl(t(\tau),r(\tau)\bigr)$,
is determined by the junction condition \cite{Israel:rt}.
 As usual, we impose the reflection symmetry with respect
to the brane. Then, we obtain the effective Friedmann equation
on the brane as \cite{Binetruy:1999ut, Maartens:2003tw}
\begin{eqnarray}
\Bigl({\dot{r}\over r} \Bigr)^2+{K\over r^2}
= \Bigl({\kappa_{5}^4\over 36}\sigma^2-{1\over l^2}\Bigr)
+{\kappa_{5}^4 \over 18}\Bigl(2\sigma\rho+\rho^2\Bigr)
+{M_{0}\over r^4},
\label{effFreidmann}
\end{eqnarray}   
where $\sigma$ and $\rho$ are the brane tension and energy density of
the matter on the brane, respectively, and $\dot{r}=dr/d\tau$
with $\tau$ being the proper time on the brane. The final term is proportional
to the mass of the bulk black hole is often called the ``dark radiation''
since it behaves as the ordinary radiation. Geometrically, it
comes from the projected Weyl tensor in the bulk, denoted commonly by
 $E_{\mu\nu}$~\cite{Shiromizu:1999wj}.
If we apply Eq.~(\ref{effFreidmann}) to the real universe, the values
of $\sigma$, $\ell$ and $M_{0}$ are constrained 
by observations of the cosmological parameters \cite{Ichiki:2002eh}.

When the bulk ceases to be pure AdS-Schwarzschild, or when there 
exists a dynamical degree of freedom other than the metric,
the parameter $M_{0}$ is no longer constant in general, but becomes
dynamical. For instance, it is the case of the so-called bulk inflaton
model~\cite{Himemoto:2000nd,Minamitsuji:2003pb,
Langlois:2003dd,Tanaka:2003eg,Koyama:2003yz},
or when the brane radiates gravitons into the bulk~\cite{Langlois:2003zb}.
In particular, in \cite{Minamitsuji:2003pb}, 
the dynamics of a bulk scalar field is investigated
in the context of the bulk inflaton model
under the assumption that the backreaction of
the scalar field on the geometry is small, and it is found that
there exists an interesting integral expression for the projected
Weyl tensor in terms of the energy-momentum tensor of the scalar field.
This suggests the existence of a local conservation law in the bulk
that directly relates the dark radiation on the brane to the dynamics in
the bulk.

In this paper, we investigate the case when there is non-trivial
dynamics in the bulk, and clarify the relation between the bulk geometry
and the dynamics of the brane.
We focus on the case of isotropic and homogeneous branes, hence assume
the existence of slicing by the maximally symmetric 3-space as
in Eq.~(\ref{adssch}). In this case, we can derive a local energy
conservation law in the bulk, in analogy with spherical symmetric
spacetimes in 4-dimensions~\cite{Kodama:vn}. Then, this conservation
law can be used to relate the brane dynamics with the geometrical
properties of the bulk, especially with the projected Weyl tensor
in the bulk.

The paper is organized as follows. 
In Sec.~II, we derive the local energy conservation law in the bulk
and discuss the general property of the bulk geometry
and cosmology on the brane.
 We show that there exists a unique generalization of
the dark radiation that is directly related to 
the local mass in the bulk.
We also find that there exists another
conserved current associated with the Weyl tensor,
as a non-linear version of what was found in \cite{Minamitsuji:2003pb}.
In a vacuum (Ricci-flat) spacetime, the local charge
for this current is found to be equivalent to the local mass.
Let us call this the Weyl charge.
The difference between the local mass and Weyl charge
is given by the linear combination of certain components of the bulk
energy-momentum 
tensor, and the projected Weyl tensor that appears in the
effective Friedmann equation on the brane is indeed
given by this Weyl charge.  Thus we have a unique decomposition
of the projected Weyl tensor term into 
the part due to the bulk mass that generalizes the dark radiation
term and the part due to the bulk energy-momentum tensor. 
In Sec.~III, as an application of the conservation law 
derived in Sec.~II, 
we consider a simple null dust model
and discuss the black hole formation in the bulk.
We identify the location of an apparent horizon
and analyze possible trajectories of the brane
in the bulk.
We show that the brane stays always outside of 
the apparent horizon of the black hole as long as
the brane is expanding.
In Sec~IV, we summarize our work and 
mention future issues.

\section{Local conservation law in a spacetime with maximally
symmetric 3-space}

In this section, we discuss the general property of a dynamical 
bulk spacetime with maximally symmetric 3-space,
and consider cosmology on the brane.
First, we derive a local conservation law in the bulk,
as a generalization of the local energy conservation law in a
spherically symmetric spacetime in 4-dimensions~\cite{Kodama:vn}.
Namely, we show that a locally conserved energy flux vector
exists in spite of the absence of a timelike Killing vector field.
This enables us to define a local mass in the bulk spacetime.
We also show that there exists a conserved current associated
with the Weyl tensor. This gives rise to a locally defined
Weyl charge. It is shown that the Weyl charge and the local mass
are closely related to each other. 

Next, we introduce the brane as a boundary of the dynamical
spacetime. The effective Friedmann equation is determined
via the junction condition and it is shown that the local mass
corresponds to the generalized dark radiation.
Finally, we show that the projected Weyl tensor on the brane is
uniquely related to the local mass. 

\subsection{Local conservation law}

We assume that the bulk allows slicing by a maximally symmetric 3-space.
Then, the bulk metric can written in the double-null form
\begin{eqnarray}
ds^2={4r_{,u}r_{,v}\over\Phi} du dv+r(u,v)^2d\Omega_{(K,3)}^2,
 \label{sasuke0}
\end{eqnarray}
where we refer to $v$ and $u$ as the advanced and retarded time coordinates,
respectively. In Appendix A, the explicit components of the connection and
curvature in an $(n+2)$-dimensional spacetime with maximally symmetric
 $n$-space are listed.

The 5-dimensional Einstein equations are given by
\begin{eqnarray}
G_{ab} +\Lambda_{5} g_{ab}=\kappa_{5}^2 T_{ab} + S_{ab} \delta(y-y_{0}),
\end{eqnarray} 
where the indices $\{a,b\}$ run from 0 to 3, and 5, and $\Lambda_{5}$ and
$\kappa_{5}^2$ are the 5-dimensional cosmological constant and
gravitational constant, respectively.
The brane is introduced as a singular hypersurface
locates at $y=y_{0}$, where $y$ denotes a Gaussian normal
coordinate in the direction of the extra dimension in the vicinity
of the brane, and $S_{ab}$ denotes the
 energy-momentum tensor on the brane. The spacetime is assumed to
be reflection symmetric with respect to the brane.

First, we consider the Einstein equations in the bulk.
They are given by
\begin{eqnarray}
&& 3{r_{,u}\over r}\Bigl(\log\Big|{r_{,v}\over\Phi}\Big|\Bigr)_{,u}=
\kappa_{5}^2 T_{uu},\quad
3{r_{,v}\over r}\Bigl(\log\Big|{r_{,u}\over\Phi}\Big|\Bigr)_{,v}= 
\kappa_{5}^2 T_{vv}, 
\nonumber\\
&& 6{r_{,u}r_{,v}\over r^2}\Bigl(1-{K\over\Phi}\Bigr)
+3{r_{,uv}\over r}= \kappa_{5}^2 T_{uv}-{2r_{,u}r_{,v}\over \Phi}\Lambda_{5},
\nonumber\\
&&\Biggl\{{r^2\Phi \over 2r_{,u}r_{,v}}\Bigl[\Bigl
(\log\Big|{r_{,u}r_{,v}\over\Phi}\Big|\Bigr)_{,uv}+4{r_{,uv} \over r}
\Bigr]-\Bigl(K-\Phi \Bigr)\Biggr\} \gamma_{ij}
=\kappa_{5}^2 T_{ij}-r^2 \gamma_{ij}\Lambda_{5},
\label{Eins}
\end{eqnarray}
where $\gamma_{ij}$ is the intrinsic metric of the maximally
symmetric 3-space.

Now, we derive the local conservation law. 
We introduce a vector field in 5-dimensional spacetime as
\begin{eqnarray}
\xi^a=\frac{1}{2}\Phi \left(-\frac{1}{r_{,v}}\frac{\partial}{\partial v}
+\frac{1}{r_{,u}}\frac{\partial}{\partial u}\right)^a\,.
\label{Kdef}
\end{eqnarray}
{}From the form of the metric (\ref{sasuke0}),
we can readily see that $\xi^a$ is conserved:
\begin{eqnarray}
\sqrt{-g}\,\xi^a{}_{;a}=\bigl(\sqrt{-g}\,\xi^a\bigr)_{,a}
=2\sqrt{\gamma}
\left((r^3r_{,u})_{,v}-(r^3r_{,v})_{,u}\right)=0\,,
\label{Kconv}
\end{eqnarray}
where $\gamma=\det\gamma_{ij}$.
Note that, for an asymptotically constant curvature spacetime,
the vector field $\xi^a$ becomes asymptotically the timelike Killing
 vector field $-(\partial/ \partial t)^{a}$.

With this vector field $\xi^a$, we define a new vector field,
\begin{eqnarray}
\tilde{S}^{a}=\xi^b\tilde{T}_{b}{}^{a},
\end{eqnarray}
where
\begin{eqnarray}
\tilde{T}_{ab}= T_{ab}-{1\over \kappa_{5}^2}\Lambda_{5}g_{ab}. 
\label{tildeT}
\end{eqnarray} 
Using the Einstein equations,
the components of the vector field $\tilde{S}^{a}$ are given by
\begin{eqnarray}
&& \kappa_{5}^2 \sqrt{-g} \tilde{S}^{v}={3\over 2}
\Bigl[ r^2\bigl(K-\Phi\bigr) \Bigr]_{,u}\sqrt{\gamma} \,,
\nonumber\\
&& \kappa_{5}^2\sqrt{-g} \tilde{S}^{u}=-{3\over 2}
\Bigl[ r^2\bigl(K-\Phi\bigr) \Bigr]_{,v}\sqrt{\gamma}\,.
\label{SVU}
\end{eqnarray}
Then, we have the local conservation law as
\begin{eqnarray}
 \tilde{S}^{a}{}_{;a}=0.
\label{conserv}
\end{eqnarray}
Since $\xi^a$ is conserved separately, the conservation of ${\tilde S}^a$
implies that we have another conserved current $S^a$ defined by
\begin{eqnarray}
S^{a}:=\xi^bT_b{}^a
\left(={\tilde S}^a+\frac{1}{\kappa_5^2}\Lambda_5\xi^a\right)\,.
\label{Sdef}
\end{eqnarray}
Thus we have the local conservation law for the energy-momentum 
tensor in the bulk.

{}From Eqs.~(\ref{SVU}), we readily see 
the local mass corresponding to $\tilde S^a$ is given
by \cite{Kodama:vn}
\begin{eqnarray}
\tilde{M}:=\Bigl(K-\Phi\Bigr)r^2,
 \label{tildeM}
\end{eqnarray}
where the factor $3/2$ in the original expression for $\tilde S^a$
is eliminated for later convenience.
Alternatively, corresponding to $S^a$, we have another local mass
that excludes the contribution of the bulk cosmological constant,
\begin{eqnarray}
M:=\tilde{M}-{1\over 6} \Lambda_{5} r^4
=\Bigl(K-\Phi\Bigr)r^2-\frac{1}{6}\Lambda_5r^4\,.
\end{eqnarray}
In what follows, we focus on the matter part $M$,
rather than on the whole mass $\tilde{M}$.
It may be noted, however, that this decomposition of
$\tilde M$ to the cosmological constant part and the matter part
is rather arbitrary, as in the case of a bulk scalar field.
Here we adopt this decomposition just for convenience.
For example, this decomposition is more useful when we
consider small
perturbations on the static AdS-Schwarzschild bulk.

We note that, in the case of a spherically symmetric
asymptotic flat spacetime in 4-dimensions
(hence $K=+1$ and with no cosmological constant),
this function $M$ agrees with the 
Arnowitt-Deser-Misner (ADM) energy or the Bondi energy
in the appropriate limits.

\subsection{Local mass and Weyl charge} 

{}From the 5-dimensional Einstein equations~(\ref{Eins}), 
we can write down the local conservation equation for
$M$ in terms of the bulk energy-momentum tensor explicitly as
\begin{eqnarray}
&&M_{,v}= {2\over 3 }\kappa_5^2r^3
 \Bigl(T^{u}{}_{v}r_{,u}-T^{v}{}_{v}r_{,v}\Bigr),
\nonumber\\    
&&M_{,u}= {2\over 3 }\kappa_5^2r^3
 \Bigl(T^{v}{}_{u}r_{,v}-T^{v}{}_{v}r_{,u}\Bigr),
\end{eqnarray}
or in a bit more concise form,
\begin{eqnarray}
dM=\frac{2}{3}\kappa_{5}^2r^3
 \Bigl(T^{u}{}_{v}r_{,u}dv+T^{v}{}_{u}r_{,v}du-T^{v}{}_{v}dr\Bigr).
\label{nulllocal} 
\end{eqnarray}  
Using the above, we can immediately write down two
integral expressions for $M$ given in terms of
flux crossing the $u=\mathrm{constant}$ hypersurfaces
from $v_{1}$ to $v_{2}$,
 and flux crossing the $v=\mathrm{constant}$ hypersurfaces
from $u_{1}$ to $u_{2}$, respectively, as
\begin{eqnarray}
&& M(v_2,u)-M(v_1,u)
= {2 \over 3 }\kappa_{5}^2 
\int^{v_2}_{v_1}\, dv\,r^3
 \Bigl(T^{u}_{\,\,\,v}r_{,u}-T^{v}_{\,\,\,v}r_{,v}\Bigr)
\Big|_{u={\rm const.}},
\nonumber\\ 
&& M(v,u_2)-M(v,u_1)
= {2 \over 3 }\kappa_{5}^2 
\int^{u_2}_{u_1}\, du\, r^3
 \Bigl(T^{v}_{\,\,\,u}r_{,v}-T^{v}_{\,\,\,v}r_{,u}\Bigr)
\Big|_{v={\rm const.}}\,.
\label{sasuke1} 
\end{eqnarray}

Finally, let us consider the Weyl tensor in the bulk.
In the present case of a 5-dimensional spacetime with
maximally symmetric 3-space, there exists only one non-trivial
component of the Weyl tensor, say $C_{vu}{}^{vu}$.
The explicit expressions for
the components of the Weyl tensor are given in Appendix A,
Eqs.~(\ref{Weylcomp}). 
Using the Bianchi identities
and the Einstein equations, we have~\cite{Wald:rg}
\begin{eqnarray}
C_{abcd}^{\quad\,\,\,\,;d}= J_{abc},
\end{eqnarray}
 where
\begin{eqnarray}
J_{abc}={2(n-1)\over n} \kappa_{n+2}^2 \Bigl(T_{c[a;b]}
+{1\over (n+1)}g_{c[b}T_{;a]}\Bigr). 
\end{eqnarray}
{}From this, we can show that there exists 
a conserved current,
\begin{eqnarray}
Q^{a}=r\,\ell_{b}n_{c}J^{bca}\,;\quad
Q^a{}_{;a}=0\,,
\end{eqnarray}
where $\ell_a$ and $n_{a}$ are a set of
two hypersurface orthogonal null vectors,
\begin{eqnarray}
\ell_a&=&\sqrt{\frac{2}{\Phi}}\left(-r_{,v}\,dv\right)_a\,,
\quad
\ell^{a}=-\sqrt{{1\over2}\Phi}\,\frac{1}{r_{,u}}
\left(\frac{\partial}{\partial u}\right)^a,
\nonumber\\
\quad
n_a&=&\sqrt{\frac{2}{\Phi}}\left(r_{,u}du\right)_a\,,
\quad
n^{a}=\sqrt{{1\over2}\Phi}\,\frac{1}{r_{,v}}
\left(\frac{\partial}{\partial v}\right)^a.
\label{ellandn}
\end{eqnarray}
The non-zero components are written explicitly as
\begin{eqnarray}
Q^{u}=-rJ^{vu}{}_{v}\,,\quad Q^{v}=-rJ^{vu}{}_{u}\,,
\end{eqnarray}
and we have
\begin{eqnarray}
&& \Bigl(r^{4} C_{vu}{}^{vu}\Bigr)_{,v}
=r^{4} J^{v}{}_{vv}\,,  \label{4dWeyl}
 \\ \nonumber 
 && \Bigl(r^{4} C_{vu}{}^{vu}\Bigr)_{,u}
=r^{4} J^{u}{}_{uu}\,.
\end{eqnarray} 
These are very similar to Eqs.~(\ref{SVU}). 
It is clear that $r^{4} C_{vu}{}^{vu}$ defines
a local charge associated with this conserved current,
that is, the Weyl charge.

Using the Einstein equations, 
we then find that the Weyl charge can be
expressed in terms of $M$ and the energy-momentum tensor as
\begin{eqnarray}
r^4 C_{vu}{}^{vu}
=3\tilde{M}+{r^4\over 6} \Bigl(6G^{v}{}_{v}-G^{i}{}_{i}\Bigr)
=3M+\frac{\kappa_5^2}{6}\,r^4 \Bigl(6T^{v}{}_{v}-T^{i}{}_{i}\Bigr).
\label{M}
\end{eqnarray} 
This is one of the most important results in this paper. 
As we shall see below, the Weyl component $C_{vu}{}^{vu}$ is
directly related to the projected Weyl tensor $E_{\mu\nu}$,
and hence this relation gives explicitly how the local mass
$M$ and the local value of the energy-momentum tensor
affects the brane dynamics.

\subsection{Apparent horizons}

As in the conventional 4-dimensional gravity, the gravitational
dynamics may lead to the formation of a black hole
in the bulk. Rigorously speaking, the black hole formation can be
discussed only by analyzing the global causal structure of
a spacetime. 
Nevertheless, 
we discuss 
the black hole formation by studying the formation of an apparent horizon.

In 4-dimensions, an apparent horizon is defined as a closed 2-sphere
on which the expansion of an outgoing (or ingoing) null geodesic
congruence vanishes. Here, we extend the definition to our case 
and define an apparent horizon as a 3-surface on which the
expansion of a radial null geodesic congruence vanishes.
Note that `radial' here means simply those congruences
that have only the $(v,u)$ components, hence an apparent
horizon will not be a closed surface if $K=0$.

The expansions of the congruence of null geodesics forming
the $u=\mathrm{constant}$ and $v=\mathrm{constant}$ hypersurfaces,
respectively, are given by~\cite{Kodama:vn}
\begin{eqnarray}
\rho_{u} = -{1\over2} u^{;a}{}_{;a} 
= -{1\over 2r} {\Phi\over r_{,u}}\,,
\quad
 \rho_{v} = -{1\over2} v^{;a}{}_{;a}
 = -{1\over 2r} {\Phi\over r_{,v}}\,.
\end{eqnarray} 
 Naively, if $\Phi=0$, one might think 
that both $\rho_{u}$ and $\rho_{v}$ vanish.
However, from the regularity condition of the metric~(\ref{sasuke0}),
we have
\begin{eqnarray}
-4{r_{,u}r_{,v}\over \Phi}>0\,.
\end{eqnarray}
Hence, it must be that $r_{,u}=0 $ or $r_{,v}=0$, if $\Phi=0$.
If $\Phi=r_{,v}=0$, we have $\rho_{u}=0$ and an
apparent horizon for the outgoing null geodesics is
formed, whereas if $\Phi=r_{,u}=0$, we have $\rho_{v}=0$ 
and an apparent horizon for the ingoing null geodesics is
formed.

\subsection{Brane cosmology}

We now consider the dynamics of a brane
in a dynamical bulk with maximally symmetric 3-space~\cite{Ida:1999ui}.
The brane trajectory
is parameterized as $(v,u)=\bigl(v(\tau),u(\tau)\bigr)$.
Taking $\tau$ to be the proper time on the brane,
we have 
\begin{eqnarray}
4{r_{,u}r_{,v}\over \Phi}\dot{u}\, \dot{v}=-1\,,
\label{sasuke4}
\end{eqnarray}
on the brane, where $\dot{u}=du/d\tau$ and so on.
The unit vector tangent to the brane (i.e., the 5-velocity of the brane) 
is given by
\begin{eqnarray}
v^{a}
=\left(\dot{v}\frac{\partial}{\partial v}
  +\dot{u}\frac{\partial}{\partial u}\right)^a,
\quad
v_{a}={2r_{,u}r_{,v}\over \Phi} 
\Bigl(\dot{u}\,dv+\dot{v}\,du\Bigr)_a\,, 
\end{eqnarray}
and the unit normal to the brane is given by
\begin{eqnarray}
n^{a}=\left(-\dot{v}\frac{\partial}{\partial v}
+\dot{u}\frac{\partial}{\partial u}\right)^a,
\quad
n_{a}={2r_{,u}r_{,v}\over \Phi}
 \Bigl(\dot{u}\,dv-\dot{v}\,du\Bigr)_a\,. 
\end{eqnarray}
The components of the induced metric on the brane are
calculated as
\begin{eqnarray}
q_{\mu\nu}={\partial x^{a}\over \partial y^{\mu}}
{ \partial x^{b}\over \partial y^{\nu}} g_{ab}\,,
\end{eqnarray}
where $\mu$, $\nu$ run from 0 to 3 and $y^{\mu}$ are
 the intrinsic coordinates on the brane with $y^0=\tau$
and $y^i=x^i$ ($i=1,2,3$).
Then the induced metric on the brane is given by 
\begin{eqnarray}
ds_{(4)}^2=-d\tau^2 +r(\tau)^2 d\Omega_{(K,3)}^2, 
\label{sasuke3}
\end{eqnarray}

The trajectory of the brane is determined by the junction
condition under the $Z_{2}$ symmetry with respect to the
brane. The extrinsic curvature on the brane is determined as
\begin{eqnarray}
 K_{\mu\nu} = -{\kappa_{5}^2 \over 2}\Bigl(S_{\mu\nu}-{1\over 3}
S q_{\mu\nu}\Bigr),
 \label{sasuke5}
\end{eqnarray}   
where $S_{\mu\nu}$ is assumed to take the form
\begin{eqnarray}
S^{\mu}_{\,\,\nu}= {\rm diag}.\bigl(-\rho,p,p,p\bigr)
-\sigma \delta^{\mu}_{\,\,\nu}\,,
\end{eqnarray}
with $\sigma$ and $\rho$ being the tension and energy
density of the matter on the brane, respectively,
as introduced previously, and $p$ being the isotropic
pressure of the matter on the brane.
Substituting the induced metric~(\ref{sasuke3}) in
 Eq.~(\ref{sasuke5}), we obtain
\begin{eqnarray}
&&
r_{,u}\dot{u}=-{r\over2}\Bigl[{\kappa_{5}^2\over 6}
\Bigl(\rho+\sigma\Bigr)-H\Bigr], 
\label{dotu}\\ 
&&
r_{,v}\dot{v}={r\over2}\Bigl[{\kappa_{5}^2\over 6}
 \Bigl(\rho+\sigma\Bigr)+H\Bigr],
\label{dotv}
\end{eqnarray}
where $H=\dot r/r$.
Multiplying the above two equations and using the normalization
condition~(\ref{sasuke4}),
we then obtain the effective Friedmann equation on the brane:
\begin{eqnarray}
H^2+{K\over r^2}=
\Bigl({\kappa_{5}^4\over 36}\sigma^2-{1\over l^2}\Bigr)
+{\kappa_{5}^4 \over 18}
\Bigl(2\sigma\rho +\rho^2\Bigr)+{M\over r^4}\,.
 \label{sasuke6}  
\end{eqnarray}
We see that $M$ is a natural generalization of the dark radiation
in the AdS-Schwarzschild case to a dynamical bulk.

For a dynamical bulk, $M$ varies in time.
The evolution of $M$ is determined by Eq.~(\ref{nulllocal}), 
and on the brane it gives
\begin{eqnarray}
\dot{M}
&=&M_{,v} \dot{v}+M_{,u} \dot{u}
\nonumber\\
&=&
{2\over 3}\kappa_{5}^2 r^4\Bigl[
T_{vv}\Bigl({1\over 6}\kappa_{5}^2
\bigl(\rho+\sigma\bigr)-H\Bigr)\dot{v}^2
-T_{uu}\Bigl({1\over 6}\kappa_{5}^2
\bigl(\rho+\sigma\bigr)+H\Bigr)\dot{u}^2
\Bigr] 
-{2\over 3}\kappa_{5}^2 r^4 H T^{v}{}_{v}\, . 
\label{varM}
\end{eqnarray}
This result is consistent with \cite{Tanaka:2003eg,Langlois:2003zb}.
{}From the Codacci equation on the brane \cite{Shiromizu:1999wj},
\begin{eqnarray}
D_{\nu}K^{\nu}{}_{\mu}-D_{\mu}K^\nu{}_\nu
= \kappa_{5}^2T_{ab}n^{b}q^{a}{}_{\mu}\,,
\label{Codacci}
\end{eqnarray}
where $D_{\mu}$ is the covariant derivative with respect to
$q_{\mu\nu}$ and $K_{\mu\nu}$ is the extrinsic curvature of the brane,
we obtain the equation for the energy transfer of the matter
on the brane to the bulk,
\begin{eqnarray}
\dot\rho+3H(\rho+p)=
2\Bigl(-T_{vv}\dot{v}^2+T_{uu}\dot{u}^2\Bigr). 
\label{energyeq}
\end{eqnarray}
Equations~(\ref{sasuke6}), (\ref{varM}) and (\ref{energyeq})
determine the cosmological evolution on the brane, once
the bulk geometry is solved.
These equations will be applied to a null dust model in the next section.
The case of the Einstein-scalar theory in the bulk is briefly discussed
in Appendix B.

Now we relate the above result with
the geometrical approach developed in~\cite{Shiromizu:1999wj},
in particular with the $E_{\mu\nu}$ term on the brane.
The projected Weyl tensor
\begin{eqnarray}
E_{\mu\nu}=C_{a\mu b \nu }\,n^{a}n^{b},  
\end{eqnarray}
has only one non-trivial component as
\begin{eqnarray}
E_{\tau\tau}= C_{abcd}n^{a}n^{c} v^{b} v^{d}
       =4\, C_{uvuv} \dot{u}^2\dot{v}^2
      =- C_{vu}{}^{vu}.
\label{EMT}
\end{eqnarray}
Using Eq.~(\ref{M}), this can be uniquely decomposed into the
part proportional to $M$ and the part due to the
projection of the bulk energy-momentum tensor on the brane.
We find
\begin{eqnarray}
E_{\tau\tau}
=-\frac{3\tilde M}{r^4}
+\frac{1}{6}\Bigl(G^i{}_i-6G^v{}_v\Bigr)
=-\frac{3M}{r^4} +\frac{\kappa_{5}^2}{6}
\Bigl(T^{i}{}_{i}-6T^{v}{}_{v}\Bigr). 
\label{Etautau}
\end{eqnarray}
If we eliminate the $M/r^4$ term from Eq.~(\ref{sasuke6}) by 
using this equation,
we recover the effective Friedmann equation on the brane
in the geometrical approach~\cite{Shiromizu:1999wj},
\begin{eqnarray}
H^2+{K\over r^2}=
\Bigl({\kappa_{5}^4\over 36}\sigma^2-{1\over l^2}\Bigr)
+{\kappa_{5}^4 \over 18}
\Bigl(2\sigma\rho +\rho^2\Bigr)
+\kappa_5^2T^{(b)}_{\tau\tau}-{E_{\tau\tau}\over 3}\,,
 \label{gFreidmann}
\end{eqnarray}
where $T^{(b)}_{\tau\tau}$ comes from 
the projection of the bulk energy-momentum tensor on the brane
and is given in the present case by
\begin{eqnarray}
T^{(b)}_{\tau\tau}={1\over6}T^i{}_i-T^v{}_v\,.
\end{eqnarray}

Finally, from the brane point of view, it may be worthwhile to give
the expressions for
the effective total energy density and pressure on the brane.
They are given by
\begin{eqnarray}
\rho^{({\rm tot})}=\rho^{({\rm brane})}+\rho^{({\rm bulk})}\,, 
\quad
\label{effective} 
 p^{({\rm tot})}= p^{({\rm brane})}+p^{({\rm bulk})}\,,
\end{eqnarray}
where
\begin{eqnarray}
&&\kappa_{4}^2 \rho^{({\rm brane})}=
3\Bigl[{1\over 6}\kappa_{5}^2\Bigl(\rho+\sigma\Bigr)\Bigr]^2 \,,
\quad
\kappa_{4}^2 p^{({\rm brane})}=
{1\over12}\kappa_{5}^4\bigl(\rho+\sigma
\bigr)\bigl(\rho-\sigma+2p\bigr)\,,
\nonumber\\
&&\kappa_{4}^2 \rho^{({\rm bulk})}={3\tilde{M}\over r^4}\,,
\quad
\kappa_{4}^2 p^{({\rm bulk})}={\tilde{M}\over r^4} 
+{1\over 3}\kappa_{5}^2\Bigl(-{\dot{u}\over \dot{v}}\tilde{T}^{v}_{\,\,\,u}
-{\dot{v}\over \dot{u}}\tilde{T}^{u}_{\,\,\,v}+2\tilde{T}^{v}_{\,\,\,v}
 \Bigr)\,,
\end{eqnarray}
where
$\tilde{M}$ is given by Eq.~(\ref{tildeM})
and $\tilde T^a{}_b$ is defined by Eq.~(\ref{tildeT}),
and both contain the contribution from the bulk cosmological constant.
It may be noted that, unlike the effective energy density,
the effective pressure contains a part coming from the bulk
that cannot be described by the local mass alone.
The contracted Bianchi identity implies 
the conservation law for the total effective
energy-momentum on the brane:
\begin{eqnarray}
\dot\rho^{({\rm bulk})}
+3H \Bigl(\rho^{({\rm bulk})}+p^{({\rm bulk})}\Bigr)
=-\dot\rho^{({\rm brane})}
-3H \Bigl(\rho^{({\rm brane})}+p^{({\rm brane})}\Bigr)\,.
 \label{Effconserve}
\end{eqnarray}
This is mathematically equivalent to Eq.~(\ref{varM}). 
However, these two equations have different interpretations.
{}From the bulk point of view, Eq.~(\ref{varM}) is more relevant,
which describes the energy exchange between the brane and
the bulk, whereas a natural interpretation of Eq.~(\ref{Effconserve})
is that it describes the energy
exchange between two different matters on the brane; the intrinsic
matter on the brane and the bulk matter induced on the brane.
The important point is, as mentioned above,
that the pressure of the bulk matter 
has contributions not only from the local mass but also from
a projection of the bulk energy-momentum tensor, which makes the
equation of state different from $p^{(\rm bulk)}=\rho^{(\rm bulk)}/3$, 
i.e., that of a simple dark radiation.

\section{Application to Null Dust Model} 

In this section, by using the local mass derived in the previous
section, we discuss the bulk geometry and brane cosmology
in the context of a null dust model.
Especially, we pay attention to the gravitational collapse due
to the emission of energy from the brane. 
Namely, we consider an ingoing null dust fluid emitted from
the brane~\cite{Langlois:2003zb, Vaidya:1953np, Nayeri:2000pv}. 

\subsection{Set-up}

The energy-momentum tensor of a null dust fluid
takes the form \cite{Poisson:eh},
\begin{eqnarray}
T_{ab}=\mu_{1} \ell_{a}\ell_{b}+\mu_{2}n_{a}n_{b},
\end{eqnarray} 
where $\ell_a$ and $n_a$ are the ingoing and outgoing null vectors,
respectively, introduced in Eqs.~(\ref{ellandn}). 
If we require that the energy-momentum conservation law is satisfied
for the ingoing and outgoing null dust independently,
we have
\begin{eqnarray}
\mu_{1}={\Phi \over (r_{,v})^2 r^3} {f(v)\over 2},
\quad
\mu_{2}={\Phi \over (r_{,u})^2 r^3} {g(u)\over 2},
\end{eqnarray} 
where $f(v)$ and $g(u)$ are arbitrary functions of $v$ and $u$,
respectively, and have the dimension  $( G_{5}\times \mathrm{mass})^{-1}$.
We assume the positive energy density,
i.e., $f(v)\geq0$ and $g(u)\geq0$. 
Thus, the non-trivial components of the energy-momentum tensor are
\begin{eqnarray}
T_{vv}={f(v)\over r^3}\,,\quad  T_{uu}={g(u)\over r^3}.
\label{em-null}
\end{eqnarray}

To satisfy the local conservation law in an infinitesimal interval
$(u,u+du)$ and $(v, v+dv)$, 
we find that the intensity functions $f(v)$ and $g(u)$
have to satisfy the relation,
\begin{eqnarray}
f(v)\Bigl({\Phi \over r_{,v}}\Bigr)_{,u}
=g(u)\Bigl({\Phi \over r_{,u}}\Bigr)_{,v}\,. 
\label{fgcond}
\end{eqnarray}
In general, if both $f(v)$ and $g(u)$ are non-zero,
it seems almost impossible to find an analytic
 solution that satisfies Eq.~(\ref{fgcond}).
Hence we choose to set either $f(v)=0$ or $g(u)=0$.
 In the following discussion, we focus on the
case that $g(u)=0$, that is, the ingoing null dust. 

\subsection{Bulk geometry of null dust collapse}

For $g(u)=0$, Eqs.~(\ref{nulllocal}) give
\begin{eqnarray}
M_{,v}={1\over3}\kappa_{5}^2 {\Phi\over r_{,v}} f(v) ,\quad
M_{,u}=0.
\end{eqnarray} 
The second equation implies $M=M(v)$.
Substituting Eq.~(\ref{em-null}) into the Einstein equations~(\ref{Eins}),
we find
\begin{eqnarray}
{\Phi\over r_{,v}}= e^{F(v)},
 \label{Phi_v} 
\end{eqnarray}
where the function $F(v)$ describes the freedom in the rescaling
off the null
coordinate $v$. This equation is consistent with Eq.~(\ref{fgcond}).
Thus, we obtain the solution as
\begin{eqnarray}
\Phi=r_{,v}e^{F(v)}=K+{r^2\over \ell^2} -{M(v) \over r^2}\,;
\quad
M(v)={1\over3}\kappa_{5}^2 \int_{v_0}^{v}\, dv\, e^{F(v)}\,f(v)
+M_0\,,
\label{PhiMsol}
\end{eqnarray}
where we have assumed that $f(v)=0$ for $v<v_0$, that is,
$v_{0}$ is the epoch at which the ingoing flux is turned on. 
For definiteness, we assume that the bulk is pure AdS
at $v<v_0$ and set $M_0=0$ in what follows.

Transforming the double-null coordinates $(v,u)$ to the half-null
coordinates $(v,r)$ as 
\begin{eqnarray}
r_{,u} du=dr-r_{,v} dv, 
\end{eqnarray} 
the solution is expressed as 
\begin{eqnarray}
&& ds^2=-4\Phi(r,v) e^{-2F(v)}dv^2
+4e^{-F(v)} dv\,dr+ r^2d\Omega_{(K,3)}^2, 
\end{eqnarray}
where $\Phi$ is given by the first of Eqs.~(\ref{PhiMsol}).
This is an ingoing Vaidya solution with a negative
cosmological constant~\cite{Vaidya:1953np,Langlois:2003zb}.
For an arbitrary intensity function $f(v)$, this is an
exact solution for the bulk geometry. Note that if we
re-scale $v$ as $dv\to d\bar{v}=e^{-F}dv$, $f(v)$ scales
as $f(v)\to\bar{f}(v)=e^{-2F}f(v)$, which manifestly shows
the invariance of the solution under this rescaling.

An apparent horizon for the outgoing radial null congruence
is located on the 3-space satisfying
\begin{eqnarray}
\Phi=r_{,v}=0,\quad \mathrm{while} \quad r_{,u}= \mathrm{finite}. 
\label{apa}
\end{eqnarray}
This gives
\begin{eqnarray}
r^2={\ell^2\over 2}\Bigl(\sqrt{K^2+4{M(v)\over \ell^2}} -K
\Bigr).
\label{aphorizon}
\end{eqnarray}
The direction of the trajectory of the apparent horizon is
given by
\begin{eqnarray}
{dr\over dv}=\frac{M_{,v}\ell^2r}{2(r^4+M\ell^2)}
=\frac{\kappa_5^2f(v)e^{F(v)}\ell^2 r}
{6(r^4+ M\ell^2)}\,.
\end{eqnarray}
Thus, for $f(v)>0$, $dr/dv$ is positive, which implies that
the trajectory of the apparent horizon is spacelike.

For the case of $K=+1$ or $K=0$, the apparent horizon
originates from $r=0$, while it originates from $r=\ell$
for $K=-1$.
A schematic view of the null dust collapse is shown in Fig.~1.
We assume that the the brane emits the ingoing flux
during a finite interval (bounded by the dashed lines in the figures)
and no naked singularity is formed.
For all the cases, the causal structures after
the onset of emission are very similar. The spacelike singularity
is formed at $r=0$, but it is hidden inside
the apparent horizon. 

\begin{figure}
\begin{center}
\includegraphics[width=7cm]{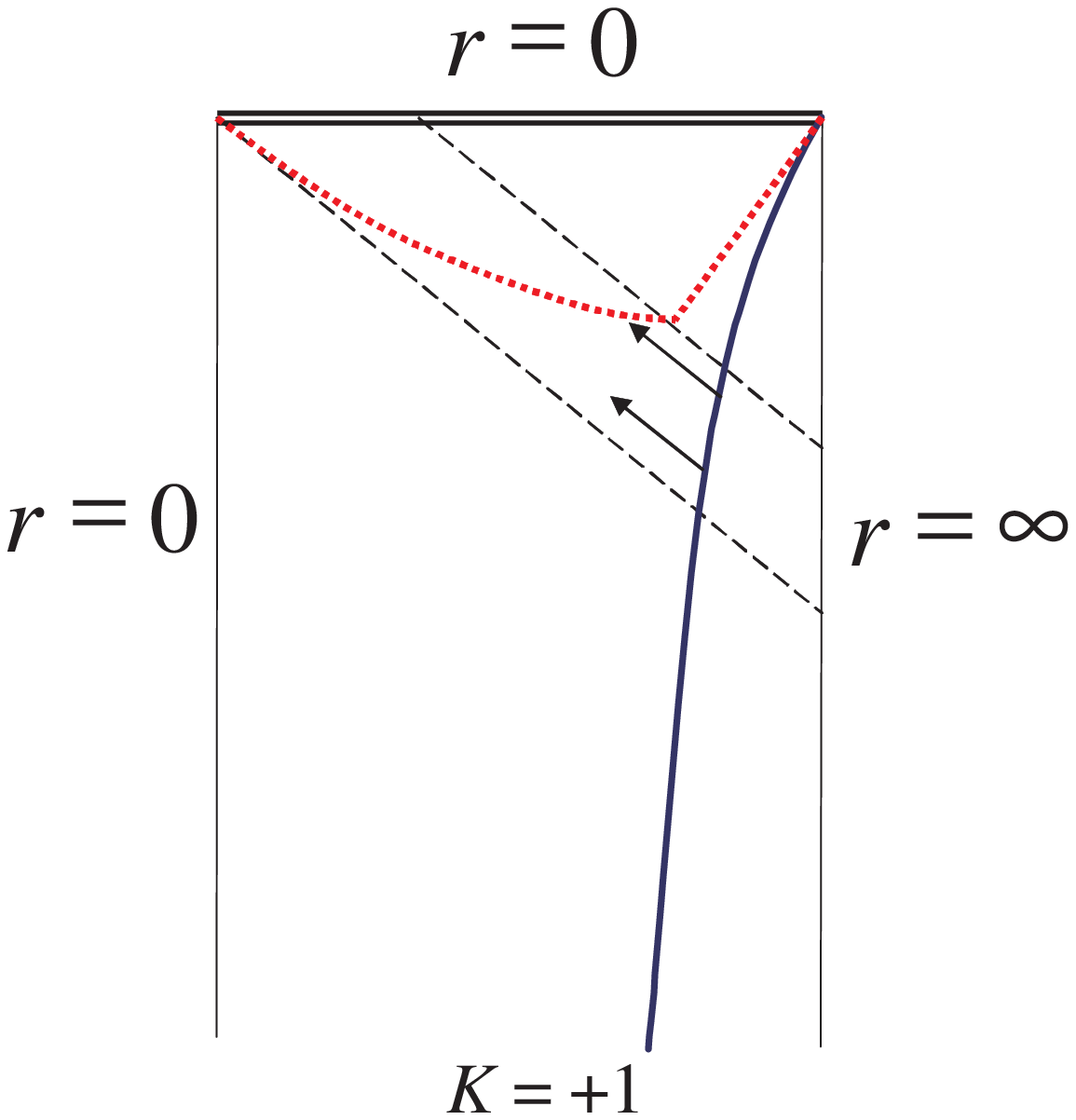}
\hspace*{10mm}\includegraphics[width=7cm]{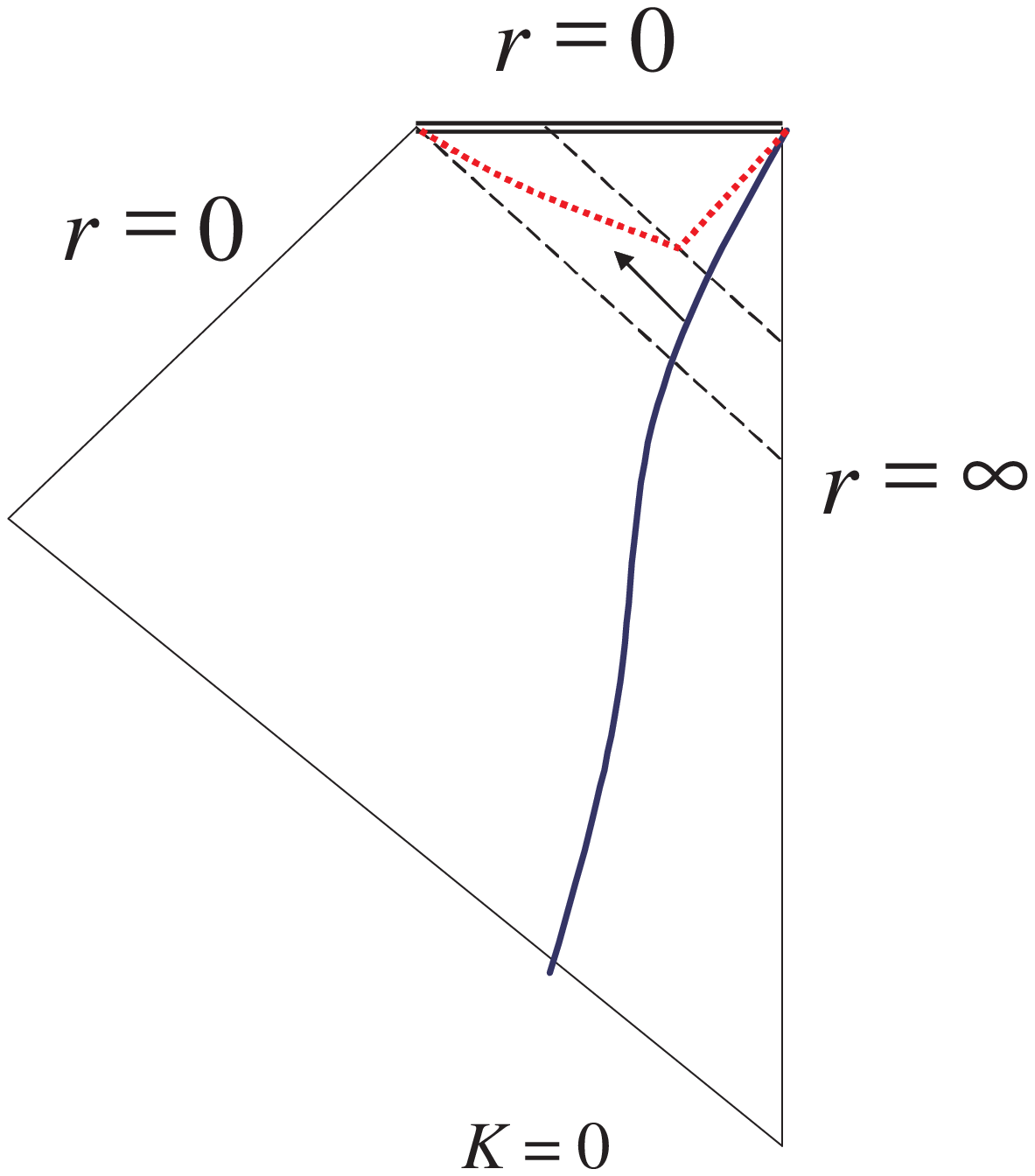}
\end{center}
\begin{center}
\includegraphics[width=10cm,height=8.5cm]{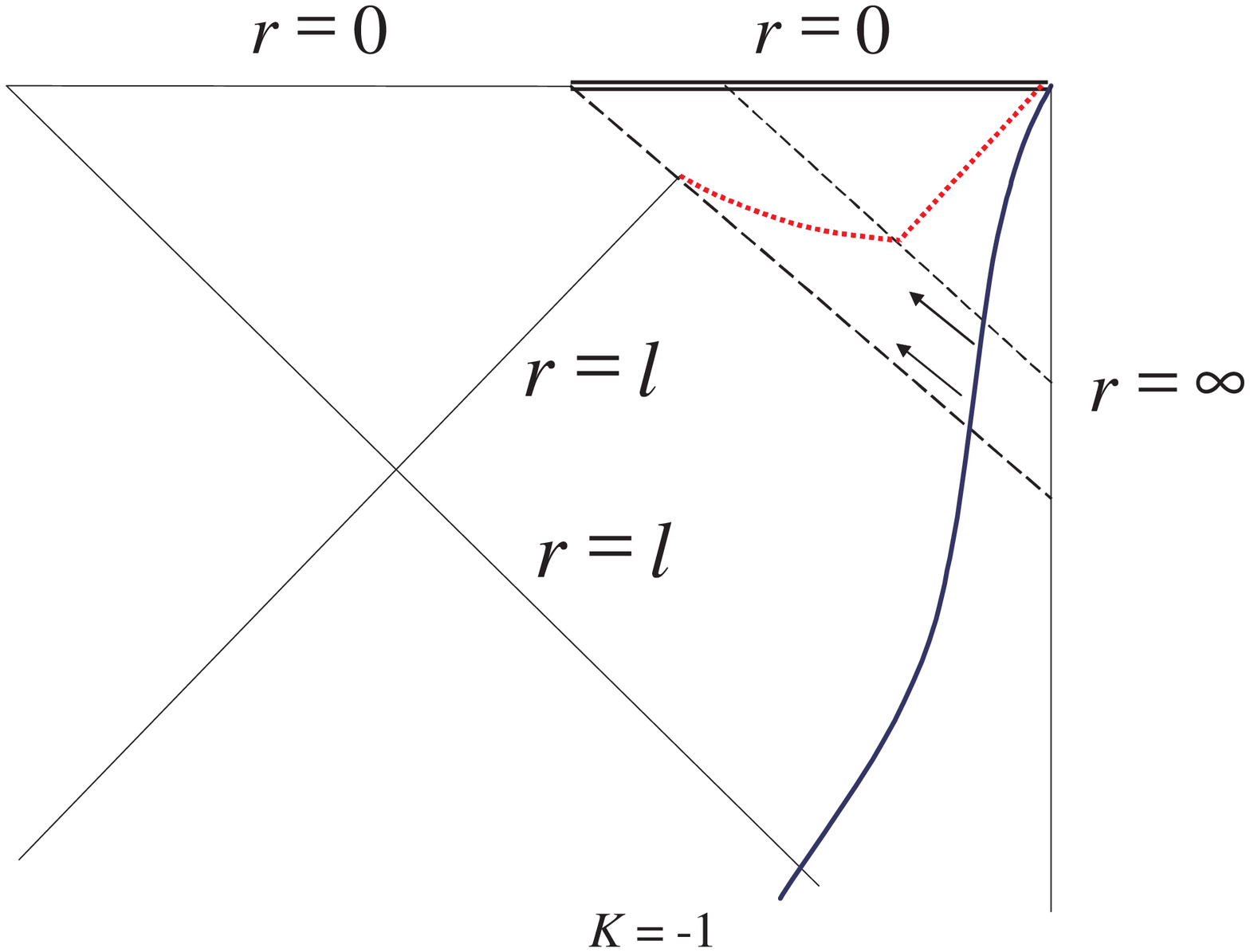}
\caption{Causal structure of a spacetime with ingoing null
dust for the cases of $K=+1$, $0$ and $-1$. In each figure,
The (almost vertical) wavy curve represents the brane trajectory
and the dotted line is the locus of the apparent horizon.
The thick horizontal line at $r=0$ 
represents the spacelike curvature singularity formed there.
The ingoing flux is assumed to be emitted during 
a finite interval bounded by the dashed lines.}
\end{center}
\end{figure}

\subsection{Brane trajectory in the bulk}

In the null dust model, using Eq.~(\ref{sasuke4}),
the proper time on the brane is related to the advanced
time in the bulk as~\cite{Nayeri:2000pv}
\begin{eqnarray}
\dot{v}_{\pm}
 =e^{F(v)}\frac{\dot{r} \pm \sqrt{\dot{r}^2+\Phi}}{2\Phi}\,.
\label{advanced}
\end{eqnarray}
To determine the appropriate sign in the above, we
require that the brane trajectory is timelike, hence $\dot{v}>0$,
and examine the signs of $\dot{v}_{\pm}$ for all possible
cases:
\begin{eqnarray*}
(1)
&&
\dot{r}>0,\ \Phi>0\quad\to\quad
\dot{v}_{+}>0,\ \dot{v}_{-}<0. 
\\
(2)
&&
\dot{r}>0,\ \Phi<0\quad\to\quad
\dot{v}_{+}<0,\ \dot{v}_{-}<0.
\\
(3)
&&
\dot{r}<0,\ \Phi>0\quad\to\quad
\dot{v}_{+}>0,\ \dot{v}_{-}<0.
\\
(4)
&&
\dot{r}<0,\ \Phi<0\quad\to\quad
\dot{v}_{+}>0,\ \dot{v}_{-}>0. 
\end{eqnarray*}
{}From these, we can conclude the following.
For an expanding brane, $\dot r>0$, the brane exists always outside
the horizon, $\Phi>0$, and $\dot v$ is given by $\dot v_{+}$.
On the other hand, a contracting brane, $\dot r<0$, can
exist either outside or inside of the horizon. Thus, 
if the brane is expanding initially, the trajectory is 
given by $\dot v=\dot v_{+}$, and it stays outside the horizon
until it starts to recollapse, if ever.
If the brane universe starts to recollapse, which is possible only
in the case $K=+1$, by continuity, the trajectory is still given by
$\dot v=\dot v_{+}$, and the brane universe is
eventually swallowed into the black hole.

{}From the above result, we find
\begin{eqnarray}
r_{,u}\dot u=\dot r-r_{,v}\dot v
=\frac{\dot r-\sqrt{\dot r^2+\Phi}}{2}<0.
\end{eqnarray}
Using Eq.~(\ref{dotu}), this gives an upper
bound of the Hubble parameter on the brane as
\begin{eqnarray}
H<{1\over 6}\kappa_{5}^2\Bigl(\rho+\sigma\Bigr). 
\label{Hubble} 
\end{eqnarray}

Let us now turn to the effective Friedmann equation on the brane.
For simplicity, we tune the brane tension to the Randall-Sundrum value,
$\kappa_5^2\sigma=6/\ell$. 
The effective Friedmann equation on the brane is
\begin{eqnarray}
H^2+{K\over r^2}={1\over 18}\kappa_{5}^4\rho\sigma
+{1\over 36}\kappa_{5}^4\rho^2+{M(\tau)\over r^4} \,,
\label{dustFeq}
\end{eqnarray} 
where $M(\tau)=M(v(\tau))$ for notational simplicity.
{}From Eq.~(\ref{energyeq}), 
the energy equation on the brane is given by
\begin{eqnarray}
\dot{\rho}+3{\dot{r}\over r}\bigl(\rho+p\bigr)
=-2\frac{f(\tau)}{r^3}\dot{v}^2,
\label{dustrhodot}
\end{eqnarray}
where $f(\tau)=f(v(\tau))$. From Eq.~(\ref{varM}),
the time derivative of $M$ is given by
\begin{eqnarray}
\dot{M}={2\over 3} r\kappa_{5}^2 
\Bigl[{1\over 6}\kappa_{5}^2\bigl(\rho+\sigma\bigr)
 -H  \Bigr]f(\tau) \dot{v}^2 .
\label{dustMdot}
\end{eqnarray}
Thus, from Eq.~(\ref{Hubble}), $M$ continues to increase on the brane.

The advanced time in the bulk is related to the proper time on the
brane by $\dot v_{+}$ in Eq. (\ref{advanced}).
Specifically, using the equality,
\begin{eqnarray}
\Phi=K+\frac{r^2}{\ell^2}-\frac{M}{r^2}
=r^2\left(\frac{\kappa_{5}^4}{36}(\rho+\sigma)^2-H^2\right),
\end{eqnarray}
on the brane, we have
\begin{eqnarray}
\dot{v}=\frac{e^{F(v)}}{2r}
\left(\frac{\kappa_5^2}{6}(\rho+\sigma)-H\right)^{-1}.
\end{eqnarray} 
Note that the product $f\,\dot v^2$ is invariant under the
rescaling of $v$. Once $f(\tau)$ is given, we can
solve the system of equations~(\ref{dustFeq}), (\ref{dustrhodot})
and (\ref{dustMdot}) self-consistently for a given initial condition,
and determine the bulk
geometry and the brane dynamics at the same time \cite{Langlois:2003zb}.
A quantitative analysis of the brane cosmology is left for future work.

\subsection{Formation of a naked singularity}

In the previous subsections, we 
assumed that there is no naked
singularity in the bulk.
However, it has been shown that a naked singularity can be formed
in the null dust collapse~\cite{Dwivedi:1989ap, Lake:1991fe, Wagh:1999ph}.
For instance, a naked singularity exists in a Vaidya spacetime
when the flux of radiation rises from zero
sufficiently slowly. We expect the same is
true in the present case.

Without loss of generality, we set $e^{F(v)}=2$.
We consider the following situation.
For $v<0$, the bulk geometry is purely AdS.
The radiative emission from the brane begins at $v=0$.
We choose the intensity function as
\begin{eqnarray}
f(v)=\frac{2\lambda}{\kappa_5^2}\, v\,,
\end{eqnarray}
where $\lambda$ is a positive constant.
This corresponds to the self-similar Vaidya spacetime if the cosmological
constant were absent~\cite{Dwivedi:1989ap}.
The brane ceases to emit radiation at $v=v_{0}$ and the bulk becomes 
a static AdS-Schwarzschild for $v>v_{0}$.
Thus the local mass is given by
\begin{eqnarray}
M(v) =
\left\{\begin{array}{ll}
   0 & (v< 0)
\\
 \vphantom{\displaystyle {G\over F}}
{2\over 3}\,\lambda\, v^2 & (0\le v \le  v_{0})
\\
 \vphantom{\displaystyle {G\over F}}
{2\over 3}\,\lambda \,v_{0}^2 & (v_0<v) .
\end{array}\right.
\end{eqnarray}

The singularity is formed at $(r,v)=(0,0)$,
and it is naked if there exists a future-directed
radial null geodesic emanating from it.
The null geodesics then form a Cauchy horizon.
The trajectory of a radial null geodesic is determined by the 
equation,
\begin{eqnarray}
{dr\over dv}= \frac{1}{2}
\left(K+{r^2(v)\over \ell^2} -{M(v)\over r^2(v)}\right)\,.
\label{nullgeode}
\end{eqnarray}
Let us analyze the above equation in the vicinity of $v=0$.
A future-directed radial null geodesic
exists if $x:=\lim\limits_{v\to 0}dr/dv$ is positive.
Using L'H$\hat{\mathrm{o}}$pital's theorem, we obtain
\begin{eqnarray}
x= \lim _{v\to0}\frac{r(v)}{v}
= \lim _{v\to0}\frac{dr}{dv}
=\frac{1}{2}\left(K-\frac{2\,\lambda}{3\,x^2}\right)\,.
\label{tangent}
\end{eqnarray}
It is clear that the above equation has no solution
when $K=0$ or $K=-1$. Hence no naked singularity is formed
for $K=0$ or $K=-1$.
Therefore, we consider the case $K=1$.
We introduce a function,
\begin{eqnarray}
Q(x)=3\,x^3-\frac{3}{2}\,x^2+\lambda\,.
\end{eqnarray}
Then, the condition for the naked singularity formation
is that $Q(x)=0$ has a solution for a positive $x$.
The function $Q(x)$ has a minimal point at 
$x=1/3$.
Therefore, the singularity is naked if 
\begin{eqnarray}
Q(1/3)=-\frac{1}{18}+ \lambda\leq0\,,
\end{eqnarray}
that is,
\begin{eqnarray}
0 < \lambda \le {1\over 18}\,. 
\label{lamdacond}
\end{eqnarray}
Thus, the bulk has a naked singularity for small
values of $\lambda$, i.e., for the flux of radiation
which rises slowly enough.

Our next interest is whether the naked singularity is local or global.
If it is globally naked, it may be visible on the brane.
To examine this, we integrate Eq.~(\ref{nullgeode}). 
In the vicinity of $v=0$, we find
\begin{eqnarray}
r_{\rm null}(v)=x_0\,v\,\left(1+b\,\frac{v^2}{\ell^2}
+\cdots\right)
\end{eqnarray}
where $x_0$ is the largest positive root of $Q(x)=0$;
\begin{eqnarray}
x_0=\frac{1}{6}
\left(1 + {\left( 1 - 36\,\lambda  + i\,
         6\,{\sqrt{2\lambda(1 - 18\,\lambda) }} \right) }^{1/3}
 + \left( 1 - 36\,\lambda  -i\,
        6\,{\sqrt{2\lambda(1 - 18\,\lambda) }} \right)^{1/3}\right)
\end{eqnarray}
and
\begin{eqnarray}
b=\frac{x_0^2}{2(5x_0-1)}\,.
\end{eqnarray}
{}From the form of $Q(x)$, we readily see that $x_0$ monotonically 
decreases from $1/2$ to $1/3$ as
$\lambda$ increases from $0$ to $1/18$,
 and hence $b$ is positive definite.
We compare this trajectory with
the trajectory of the apparent horizon.
It is given by Eq.~(\ref{aphorizon}) with $K=+1$.
In the vicinity of $v=0$, it gives
\begin{eqnarray}
r_{\rm app}(v)=\sqrt{\frac{2\lambda}{3}}\,v
\left(1-\frac{\lambda}{8}\frac{v^2}{\ell^2}+\cdots\right)\,.
\end{eqnarray}
Since $x_0>\sqrt{2\lambda/3}$ for all the values of $\lambda$ in the
range $0<\lambda\leq1/18$, and $dr_{\rm app}/dv$ is a decreasing
function of $v$ while $dr_{\rm null}/dv$ is an increasing function
of $v$, it follows that the null geodesic lies in the exterior
of the apparent horizon and the difference in the radius at the
same $v$ increases as $v$ increases, at least when $v$ is small.
This suggests that the singularity is globally naked.

In Fig.~2, we plot the loci of the null geodesic and the apparent
horizon.
The result is clear. The null geodesic always stays outside of the
apparent horizon, thus outside of the final event horizon at $v=v_{0}$.
Mathematically, this is due to the cosmological constant term in
Eq.~(\ref{nullgeode}), which strongly drives the null geodesic 
trajectory to larger values of $r$.
Thus, we conclude that the naked singularity is global and visible
on the brane. The causal structure in this case is illustrated in
Fig.~3.
Investigations on the effect of the visible singularity
on the brane are necessary, but they are left for future work.           

\begin{figure}
\begin{center}
\includegraphics[height=7.0cm]{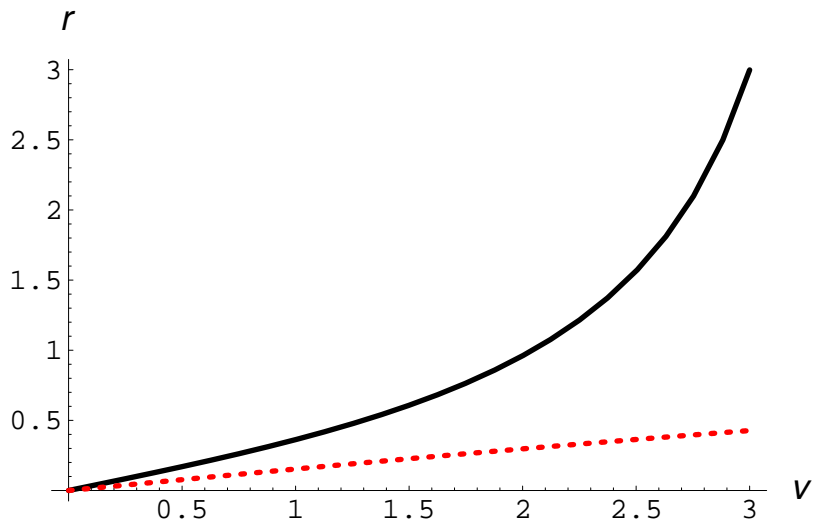}
\end{center}
\begin{center}
\caption{The loci of the null geodesic 
(the solid curve) and the apparent horizon 
(the dotted curve) on the $(v,r)$-plane, scaled
in units of the AdS radius $\ell$,
in the critical case $\lambda=1/18$. 
Their behaviors are qualitatively the same for all the
other values of $\lambda$ in the range $0<\lambda<1/18$.}
\end{center}
\end{figure}

\begin{figure}
\begin{center}
\includegraphics[width=9.0cm,height=9.0cm]{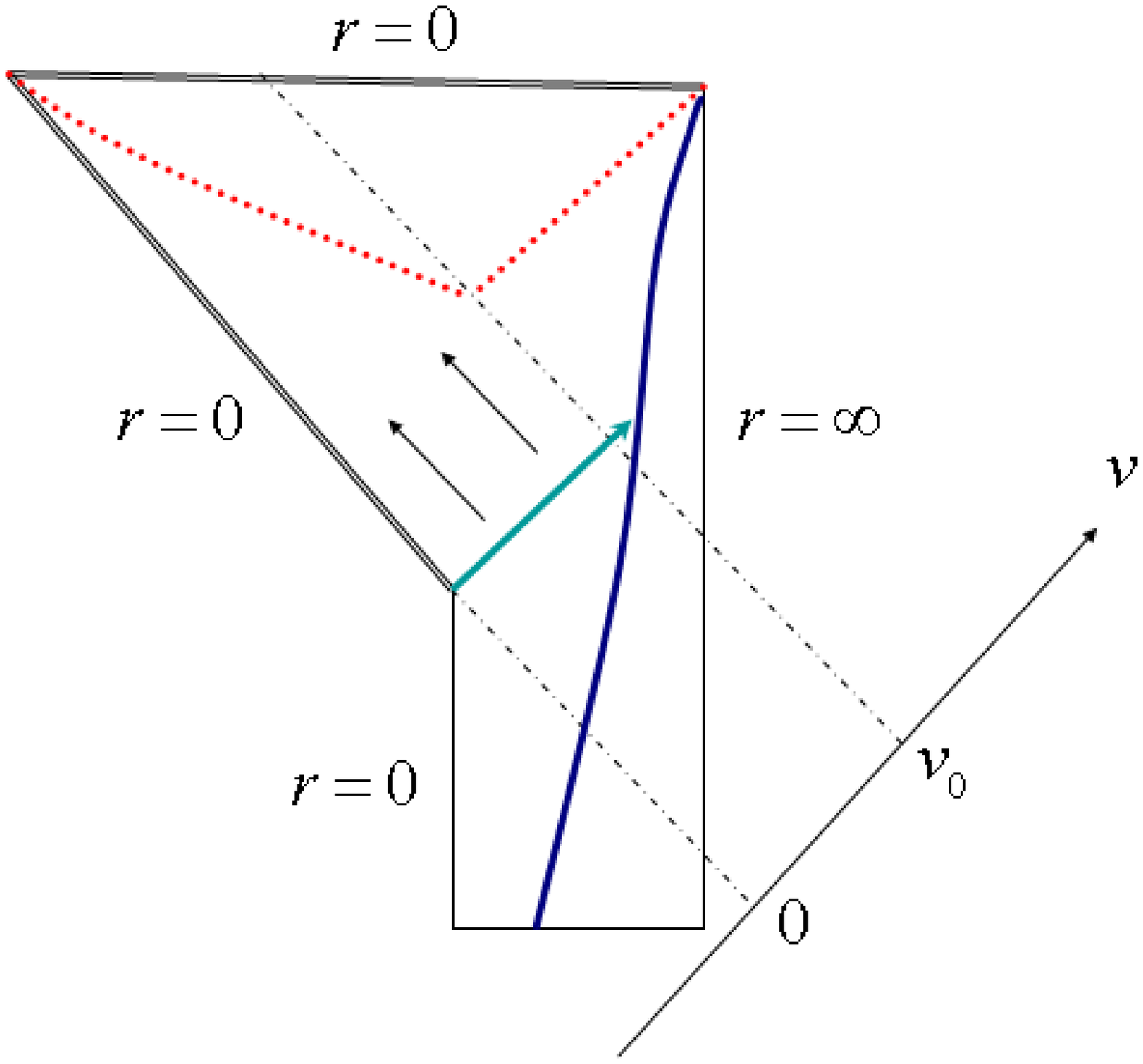}
\caption{Causal structure of a spacetime with ingoing null
dust when a naked singularity is formed.
The wavy and almost vertical curve represents the brane trajectory
and the dotted line is the locus of the apparent horizon.
A naked singularity is formed at $r=0$ along the $v=0$ null line.
A radial, future directed null geodesic originating from the naked
singularity (the right-pointed thick line) stays outside of the 
apparent horizon and reaches the brane. } 
\end{center}
\end{figure}

Finally, let us mention the strength of the naked singularity
as we approach it along a radial null geodesic. 
Let $w$ be an affine
parameter of the geodesic, $w=0$ be the singularity, and
the tangent vector be denoted by $k^{a}=dx^{a}/dw$.
We examine $R_{ab}k^a k^b$ and $C_{vu}{}^{vu}$.
{}From Eq.~(\ref{em-null}) and the Einstein equations, we have
\begin{eqnarray}
R_{ab}k^ak^b=\frac{\kappa_5^2f(v)}{r^3}\left(\frac{dv}{dw}\right)^2
=\frac{2\lambda v}{r^3}\left(\frac{dv}{dw}\right)^2
\mathop{\longrightarrow}\limits_{w\to0}\,
\frac{2\lambda}{x_{0} (1-x_{0})^2 }w^{-2}\,.
\end{eqnarray}
Also, from Eq.~(\ref{M}), we have
\begin{eqnarray}
C_{vu}{}^{vu}=\frac{3M}{r^4}=\frac{2\lambda\,v^2}{r^4}
\mathop{\longrightarrow}\limits_{w\to0}\,
\frac{2\lambda}{x_{0}^4}v^{-2} \propto w^{-\frac{2x_{0}}{1-x_{0}}}\,.
\end{eqnarray}
Thus the Ricci tensor and the Weyl tensor
diverge as $w^{-2}$ and $w^{-{2x_{0}\over 1-x_{0}}}$, respectively,
which is a sign of a strong curvature singularity.

\section{Conclusion}

In this paper, in the context of the RS2 type braneworld,
we discussed the dynamics of the bulk and the effective
cosmology on the brane in terms of the local
conservation law that exists in the bulk spacetime with maximally
symmetric 3-space.

First, we formulated the local conservation law in the dynamical
bulk.~We found that the bulk geometry is completely described by
the local mass $M$ and it is directly related to
the generalized dark radiation term in the effective Friedmann
equation. We also found that there exists a conserved current
associated with the Weyl tensor and the projected Weyl tensor 
that appears in the geometrical approach is just the local
charge for this current, and it can be expressed in terms of 
$M$ and a certain linear combination of the components of the bulk
energy-momentum tensor.

Next, as an application of our formalism, we 
adopted a simple null dust model, in which the energy emitted by
the brane is approximated by an ingoing null dust fluid,
and investigated the general properties of the bulk geometry
and the brane trajectory in the bulk. 
Usually, the ingoing null dust forms a black hole in the bulk.
However, in the case of $K=+1$, a naked singularity can be formed
in the bulk when the flux rises from zero slower than a critical rate.
We show that the naked singularity is global and thus it can be
visible to an observer on the brane. Studies on the implications of a
visible naked singularity on the brane is left for future work.

Also, we found that the brane can 
never enter the black hole horizon as long as it is expanding.
In addition, we found an upper bound on the Hubble expansion rate,
given by the energy density of the matter on the brane,
for arbitrary but non-negative energy flux emitted by the
brane. We also presented a set of equations
that completely determine the brane dynamics as well
as the bulk geometry.

Finally, let us briefly comment on some future issues.
In this paper, we only discussed the case of null dust.
However, this is too simplified to be realistic.
As a realistic situation, it will be interesting to
consider a bulk scalar field such as a dilaton or
a moduli field. In this case, it will be
necessary to solve the bulk and brane dynamics numerically
in general. Another interesting issue will be 
the evaporation of a bulk black hole by the Hawking radiation
and its effect on the brane dynamics.
We plan to come back to these issues in future publications.

\section*{Acknowledgments}
We would like to thank all the participants of the workshop
``Mini Workshop on Brane World'', held in October 2003 at Tokyo
Institute of Technology, for useful discussions.
We also would like to thank T.~Harada for reminding us
of the possibility of naked singularity formation 
in Vaidya spacetimes.
This work was supported in part by Monbu-Kagakusho
 Grant-in-Aid for Scientific Research (S) No.\ 14102004.

\appendix

\section{Geometrical quantities and local conservation laws
in $\bm{(n+2)}$-dimensions}

In this Appendix, we give useful formulas in an $(n+2)$-dimensional
spacetime with constant curvature $n$-space, and generalize the
expression for the local mass and Weyl charge.

We consider the metric in the double-null form,
\begin{eqnarray}
ds^2={4r_{,u}r_{,v}\over\Phi} du dv+r(u,v)^2d\Omega_{(K,n)}^2, 
\end{eqnarray}
where $K=+1,0$, or $-1$, corresponding to the sphere, flat space and
hyperboloid, respectively. We denote the metric tensor of the constant
curvature space as $\gamma_{ij}$.
The explicit expressions for the
geometrical quantities in this spacetime are as follows.
\begin{itemize}

\item{\bf Christoffel symbol}

\begin{eqnarray}
&& \Gamma^{u}_{uu}=\Bigl(\log\Big| {r_{,u}r_{,v}\over\Phi}\Big|\Bigr)_{,u}\,,
\quad
 \Gamma^{v}_{vv}=\Bigl(\log\Big|{r_{,u}r_{,v}\over\Phi}\Big|\Bigr)_{,v}\,,
\nonumber\\
&& \Gamma^{u}_{ij}=-{r\,\Phi\over 2r_{,u}}\gamma_{ij}\,,
\quad
   \Gamma^{v}_{ij}=-{r\,\Phi\over 2r_{,v}}\gamma_{ij}\,,
\nonumber\\
&& \Gamma^{i}_{uj}={r_{,u}\over r}\delta_{ij}\,,
\quad
 \Gamma^{i}_{vj}={r_{,v}\over r}\delta_{ij}\,,
\quad
 \Gamma^{i}_{jk}={}^{n} \Gamma^{i}_{jk}\,.
\end{eqnarray}

\item{\bf Riemann tensor}

\begin{eqnarray}
&& R^{u}{}_{uuv}=R^{v}_{\,\,vvu}=
-\Bigl(\log\Big| {r_{,u}r_{,v}\over\Phi}\Big|\Bigr)_{,uv}\,,
\nonumber \\
&& R^{u}{}_{iuj}=
\Bigl[-{1\over 2} r\Bigl({\Phi \over r_{,u}}\Bigr)_{,u}
-{r\,\Phi \over 2 r_{,u}}
\Bigl(\log\Big|{r_{,u}\,r_{,v}\over\Phi}\Big|\Bigr)_{,u} \Bigr]
\gamma_{ij}\,,
\nonumber\\
&& R^{v}{}_{ivj}=
\Bigl[-{1\over 2}r \Bigl({\Phi \over r_{,v}}\Bigr)_{,v}
-{r\,\Phi \over 2 r_{,v}}
\Bigl(\log\Big|{r_{,u}\,r_{,v}\over\Phi}\Big|\Bigr)_{,v} \Bigr]
\gamma_{ij}\,,
\nonumber\\  
&& R^{i}{}_{uju}=
\Bigl[-{r_{,uu}\over r} 
+{r_{,u} \over r}\Bigr(\log\Big|{r_{,u}r_{,v}\over\Phi}\Big|\Bigr)_{,u} \Bigr]
\delta^{i}_{j}\,,
\nonumber \\
&& R^{i}{}_{vjv}=
\Bigl[-{r_{,vv}\over r} 
+{r_{,v} \over r}\Bigr(\log\Big|{r_{,u}r_{,v}\over\Phi}\Big|\Bigr)_{,v} \Bigr]
\delta^{i}_{j}\,,
\nonumber  \\
&&R^{i}{}_{vju}=R^{i}{}_{ujv}=-{r_{,uv}\over r} \delta^{i}_{j},
\quad
R^{i}{}_{jkl}=\Bigl(K-\Phi\Bigr)
\Bigl(\delta^{i}_{k}\gamma_{jl}-\delta^{i}_{l}\gamma_{jk}\Bigr). 
\end{eqnarray}

\item{\bf Ricci tensor}

\begin{eqnarray}
&& R_{uu}
=n{r_{,u}\over r}\Bigl(\log\Big|{r_{,v}\over\Phi}\Big|\Bigr)_{,u}\,,
\quad
 R_{vv}
=n{r_{,v}\over r}\Bigl(\log\Big|{r_{,u}\over\Phi}\Big|\Bigr)_{,v} \,,
\nonumber \\
&& R_{uv}
=-\Bigl(\log\Big|{r_{,u}r_{,v}\over\Phi}\Big|\Bigr)_{,uv}
-n{r_{,uv}\over r}\,,
\nonumber\\
&& R_{ij}
=\Bigl[-{r r_{,uv}\over
r_{,u}r_{,v}}\Phi+2\bigl(n-1\bigr)\Bigl(K-\Phi\Bigr)\Bigr] 
\gamma_{ij} \,.
\end{eqnarray}

\item{\bf scalar curvature}

\begin{eqnarray}
&& R=-{\Phi \over r_{,u}r_{,v}}
\Bigl(\log\Big|{r_{,u}r_{,v}\over\Phi}\Big|\Bigr)_{,uv}
-2n{\Phi r_{,uv}\over rr_{,u}r_{,v}}+{n(n-1)\over r^2}\bigl(K-\Phi\bigr)\,.
\end{eqnarray}

\item{\bf Einstein tensor} 

\begin{eqnarray}
&& G_{uu}=R_{uu}\,,\,G_{vv}=R_{vv}\,,
\nonumber\\
&& G_{uv}=n(n-1){r_{,u}r_{,v}\over r^2}\Bigl(1-{K\over\Phi}\Bigr)
+n{r_{,uv}\over r}\,,
\nonumber\\
&& G_{ij}
=\Biggl\{{r^2\Phi \over 2r_{,u}r_{,v}}\Bigl[\Bigl
(\log\Big|{r_{,u}r_{,v}\over\Phi}\Big|\Bigr)_{,uv}+2(n-1){r_{,uv} \over r}
\Bigr]
-{(n-2)(n-1)\over 2}\Bigl(K-\Phi \Bigr)\Biggr\}\gamma_{ij}\,.
\end{eqnarray} 

\item{\bf Weyl tensor} 

\begin{eqnarray}
&&C_{uvu}{}^{u}={n-1\over n+1}
 \Bigl(\log \Big|{r_{,u}r_{,v}\over\Phi}\Big|\Bigr)_{,uv}
- {r_{,uv}\over r} 
- {r_{,u}r_{,v}\over r^2 \Phi}\bigl(K-\Phi\bigr),
\nonumber\\
&&C_{iujv}={1\over n}r^2\gamma_{ij}C_{uvu}{}^{u}
\nonumber\\
&&C_{ijkl}=-{2\over n(n-1)} r^4
 \Bigl(\gamma_{ik}\gamma_{jl}-\gamma_{il}\gamma_{jk}
 \Bigr)C_{uv}{}^{vu}.
\label{Weylcomp}
\end{eqnarray} 
 
\end{itemize}

{}From these formulas,
 we can show the existence of a conserved current in the same
way as given in the text. Namely, with the timelike vector field 
$\xi^a$ defined by Eq.~(\ref{Kdef}),
the currents $\tilde S^a=\xi^b\tilde T_b{}^a$
and $S^a=\xi^b T_b{}^a$ are separately conserved, and the corresponding
local masses are given, respectively, by 
\begin{eqnarray}
\tilde{M}=r^{n-1}\Bigl(K-\Phi\Bigr),
\end{eqnarray}
and
\begin{eqnarray} 
M=\tilde{M}-{2\over (n-1)(n-2)}\Lambda_{n+2}r^{n-1}.   
\end{eqnarray}
The $v$ and $u$ derivatives of $M$ are given by
the energy-momentum tensor as
\begin{eqnarray}
&&M_{,v}=\kappa_{n+2}^2 {2 r^n \over n}
 \Bigl(T^{u}_{\,\,\,v}r_{,u}-T^{v}_{\,\,\,v}r_{,v}\Bigr), \nonumber  \\   
&&M_{,u}=\kappa_{n+2}^2 {2 r^n  \over n}
 \Bigl(T^{v}_{\,\,\,u}r_{,v}-T^{v}_{\,\,\,v}r_{,u}\Bigr). 
\end{eqnarray}

Let us now turn to the conserved current associated with
the Weyl tensor.
We start from the equation that results from the Bianchi
 identities~\cite{Wald:rg},
\begin{eqnarray}
C_{abcd}^{\quad\,\,\,\,;d}= J_{abc}\,,
\end{eqnarray}       
 where
\begin{eqnarray}
J_{abc}={2(n-1)\over n} \kappa_{n+2}^2 \Bigl(T_{c[a;b]}
+{1\over (n+1)}g_{c[b}T_{;a]}\Bigr).
\end{eqnarray}
{}From this equation, we can 
show the existence of a locally conserved current
$Q^{a}$ given by 
\begin{eqnarray}
Q^{a}=r\,\ell_{b}n_{c}J^{bca},\quad
Q^a{}_{;a}=0\,,
\end{eqnarray}
where $\ell^{a}$ and $n^{a}$ are the null vectors defined in
Eqs.~(\ref{ellandn}).
The non-zero components are explicitly written as
\begin{eqnarray}
Q^{u}=-rJ^{vu}{}_{v}\,,\quad Q^{v}=-rJ^{vu}{}_{u}\,.
\end{eqnarray}
We then find the following relations,
\begin{eqnarray}
&& \Bigl(r^{n+1} C_{vu}{}^{vu}\Bigr)_{,v}
=r^{n+1} J^{v}{}_{vv}\,, 
 \\ \nonumber 
 && \Bigl(r^{n+1} C_{vu}{}^{vu}\Bigr)_{,u}
=r^{n+1} J^{u}{}_{uu}\,.
\end{eqnarray} 
These relations are generalization of Eqs.~(\ref{4dWeyl}),
and imply that the Weyl component $r^{n+1} C_{vu}^{\quad vu}$ 
is the local charge associated with this conserved current.

Using the explicit form of $C_{vu}{}^{vu}$ in Eqs.~(\ref{Weylcomp})
and the Einstein equations, we can relate the Weyl charge to
the local mass. We find
\begin{eqnarray}
r^{n+1}C_{vu}^{\quad vu}
&=&{n(n-1)\tilde{M}\over 2}
-{n-1\over n(n+1)} r^{n+1}\Bigl(G^{i}{}_{i}-2nG^{v}{}_{v} \Bigr) 
\nonumber\\
&=&{n(n-1)  M \over  2}
-{n-1\over n(n+1)}\, \kappa_{n+2}^2\,r^{n+1}
 \Bigl(T^{i}{}_{i}-2nT^{v}{}_{v} \Bigr). 
\label{ndWeyl}
\end{eqnarray}
Finally, we note that this equation implies that
the linear combination of the energy-momentum tensor,
\begin{eqnarray}
r^{n+1}\Bigl(T^{i}{}_{i}-2nT^{v}{}_{v}\Bigr),
\end{eqnarray}
plays the role of a local charge as well. Therefore, the behavior
of this quantity is constrained non-locally by the
integral of the flux given by the corresponding linear
combination of the currents $S^a$ and $Q^a$. 
Although we do not explore it here, this fact may be
useful in an analysis of the behavior of
the bulk matter.

\section{Einstein-scalar theory in the bulk}

In this Appendix, we apply the local conservation law to the
5-dimensional Einstein-scalar theory. We assume that there is no
matter on the brane, but we take account of a coupling of 
the bulk scalar field to the brane tension.
In this case, the energy exchange between the brane and the bulk,
hence the time evolution of $M$, occurs through the coupling.

We first consider a general bulk scalar field. Then, as
a special case, we analyze the local mass on the brane
for the exact dilatonic solution discussed
by Koyama and Takahashi~\cite{Koyama:2003yz}.
Finally, we clarify the relation between the local mass
and the term which is identified as the dark radiation term
in the effective 4-dimensional approach in which the contribution
of the scalar field energy-momentum to the brane is required
to take the standard 4-dimensional form~\cite{Langlois:2003dd}.

\subsection{Set-up}

We consider a theory described by the action,
\begin{eqnarray}
S=\int d^5 x \sqrt{-g} \Bigl[{1\over
2\kappa_{5}^2} R
-{1\over2}\partial_{a}\phi\, \partial^{a}\phi -V(\phi)\Bigr]
 -\int d^4 x  \sqrt{-q} \,\sigma(\phi)\,.
\label{bulkscalar}
\end{eqnarray} 
For the bulk with the metric given by Eq.~(\ref{sasuke0}), the
energy-momentum tensor in the bulk is given by
\begin{eqnarray}
&&T_{vv}=\phi_{,v}^2,\quad T_{uu}= \phi_{,u}^2 ,
\nonumber\\
&&T_{uv}=-{2\,r_{,u}r_{,v}\over \Phi}V(\phi), 
\nonumber\\
&&T_{ij}=-r^2\Bigl[{\Phi\over 2\,r_{,u}r_{,v}}
\phi_{,u}\phi_{,v}+V(\phi)\Bigr]\gamma_{ij}\,.
\end{eqnarray} 

On the brane, the first derivatives of the scalar field 
tangent and normal to the brane are expressed, respectively, as
\begin{eqnarray}
&&\phi':= \phi_{,a}n^{a}=-\phi_{,v}\dot{v}+\phi_{,u}\dot{u}, 
\nonumber\\
&&\dot{\phi}:= \phi_{,a}v^{a}=\phi_{,v}\dot{v}+\phi_{,u}\dot{u}. 
\end{eqnarray}
The Codacci equation~(\ref{Codacci}) gives, via the
coupling to the brane tension, the boundary condition at the brane,
\begin{eqnarray}
\phi'={1\over 2}{d\over d\phi}\sigma(\phi).\label{phiCodacci}
\end{eqnarray}
In the present case, the effective Friedmann equation induced on the brane,
Eq.~(\ref{sasuke6}), becomes
\begin{eqnarray}
&&3\Bigl[ H^2+{K\over r^2} \Bigr]
= {1\over 12}\kappa_{5}^2 \sigma^2 +{3M\over r^4}. \label{scalar-Friedmann}
\end{eqnarray}
The time evolution of the local mass $M$ on the brane is
given by
\begin{eqnarray}
\dot{M}=-{1\over3} 
\kappa_{5}^2r^4 H\Bigl[\dot{\phi}^2 -2\, V
+{1\over 4}\Bigl({d\over d\phi}\sigma\Bigr)^2\Bigr]
-{1\over36}\kappa_{5}^4r^4\dot{\phi}\,{d\over d\phi}\sigma^2.
\label{mdot}
\end{eqnarray}

{}From the brane point of view, as given by Eqs.~(\ref{effective})
in the text, the effective energy density and
pressure are composed of the brane tension and the bulk matter 
induced on the brane as
\begin{eqnarray}
\rho^{({\rm tot})}=\rho^{({\rm T})}+\rho^{({\rm B})}\,,
\quad
 p^{({\rm tot})}=p^{({\rm T})}+p^{({\rm B})}\,,
\end{eqnarray}
where 
\begin{eqnarray}
&&\kappa_{4}^2\, \rho^{({\rm T})}={1\over 12}\kappa_{5}^4 \sigma^2\,,
\quad 
\kappa_{4}^2\, p^{({\rm T})}= -{1\over 12}\kappa_{5}^4 \sigma^2,   
\nonumber\\ 
&&\kappa_{4}^2\, \rho^{({\rm B})}={3M \over r^4}\,,
\quad
\kappa_{4}^2\, p^{({\rm B})}= {M\over r^4}
+{1\over 3}\kappa_{5}^2\Bigl[\dot{\phi}^2-2V+{1\over 4}
\Bigl({d\over d\phi}\sigma\Bigr)^2\Bigr]\,,
\end{eqnarray} 
where Eq.~(\ref{phiCodacci}) is used.
{}From the Bianchi identity on the brane,
the conservation law for the total effective
energy-momentum on the brane is obtained as
\begin{eqnarray}
\dot\rho^{({\rm B})}
+3H \Bigl(\rho^{({\rm B})}+p^{({\rm B})}\Bigr)
=-\dot\rho^{({\rm T})}\,.
\label{effconserve}
\end{eqnarray}
The above relation is mathematically equivalent to Eq. (\ref{mdot}). 

As discussed after Eq.~(\ref{Effconserve}) in the text, 
Eq.~(\ref{effconserve}) gives the point of view from the brane,
and it is naturally interpreted as the equation describing the
energy exchange between the brane tension and the bulk matter
induced on the brane. On the other hand, the time variation of the
local mass along the brane, Eq.~(\ref{mdot}), gives the point of
view from the bulk, and it contains not only the energy transfer
from the brane tension to the bulk (the last term)
but also the energy flow of the bulk scalar field at the location
of the brane, which is non-vanishing in general
even if the scalar field has no coupling to
the brane tension.

\subsection{Dilatonic exact solution}

In the case $K=0$, and for special forms of $V(\phi)$ and $\sigma(\phi)$,
an exact cosmological solution is known,
as a realization of the bulk inflaton model~\cite{Koyama:2003yz}.
The forms of the potential and brane tension are
\begin{eqnarray}
&& \kappa_{5}^2 V(\phi)=\Bigl({\Delta\over 8}+\delta\Bigr) \lambda_{0}^2
e^{-2\sqrt{2} b\kappa_{5} \phi}, 
\label{dilatonpot}
\\
&& \kappa_{5}^2\sigma (\phi)= \sqrt{2} \lambda_{0}
e^{-\sqrt{2} b\kappa_{5} \phi},
\end{eqnarray}
where $\delta$, $b$ and $\lambda_{0}$ are constant and are all
assumed to be non-negative, and 
\begin{eqnarray}
\Delta= 4b^2-{8\over3}\,.
\end{eqnarray}
If $\delta=0$, there exists a static, Minkowski brane
 solution~\cite{Cvetic:2000pn}.
In order to avoid the presence of a naked singularity,
the dilatonic coupling $b^2$ is assumed to be smaller than
$1/6$~\cite{Koyama:2003yz}. This implies that $\Delta$ is
negative and it is in the range,
\begin{eqnarray}
2\leq(-\Delta)\leq\frac{8}{3}\,.
\end{eqnarray}

The exact solution takes the form
\begin{eqnarray}
ds^2
&=& e^{2W(z)} \Bigl( -d\tau^2+e^{2\alpha(\tau)} \delta_{ij}dx^i dx^j
+e^{2\sqrt{2} b\kappa_{5} \phi(\tau)} dz^2\Bigr), 
\nonumber\\ 
\phi&=&\phi(\tau) + \Xi(z),
\end{eqnarray}
with the brane located at $z=z_0$ and it is assumed that
$\Xi(z_0)=0$ without loss of generality.
The scale factor of the brane and the scalar field on the brane
are given by
\begin{eqnarray}
r(\tau)= e^{\alpha(\tau)}=\Bigl(H_{0}\tau\Bigr)^{1\over 6b^2}, 
\qquad
e^{\sqrt{2} b\kappa_{5}\phi(\tau)}= H_{0}\tau, 
\label{tau}
\end{eqnarray}
where
\begin{eqnarray}
H_{0}=\Bigl(\Delta+{8\over 3}\Bigr)\lambda_{0}\sqrt{{\delta\over -\Delta}}
=4b^2 \lambda_{0}\sqrt{{\delta\over -\Delta}}\,.
\end{eqnarray}
As seen from the first of Eqs.~(\ref{tau}), the power-law inflation
is realized on the brane for $b^2<1/6$.

Let us consider the time evolution of the energy content in this model.
{}From the brane point of view, the time derivative of the brane tension
$\rho^{({\rm T})}$ is always negative;
\begin{eqnarray}
\dot\rho^{({\rm T})}=\frac{\Delta}{48b^4 \delta\,\tau^3}<0.  
\end{eqnarray}
Thus, from Eq.~(\ref{effconserve}), for an observer on the brane,
there is one-way energy transfer from the brane tension 
to the bulk matter induced on the brane.
{}From the bulk point of view, however, the situation is
slightly more complicated.
The time derivative of the local mass (or the generalized
dark radiation) on the brane, Eq.~(\ref{mdot}),
is evaluated as
\begin{eqnarray}
{\dot{M}\over r^4}
= {1\over 18 b^6 \delta\,\tau^3}\Bigl({1\over 3}-b^2 \Bigr)
\Bigl({\Delta \over 8 } +\delta \Bigr). 
\end{eqnarray}
The sign of $\dot M$ is determined by the sign of $\Delta/8+\delta$.
Note that the sign of $\Delta/8+\delta$ determines the sign
of the bulk potential as well, as seen from Eq.~(\ref{dilatonpot}).
If $\Delta/8+\delta>0$, i.e., $\delta>(-\Delta)/8=(b^2/ 2)-1/3$,
we have $\dot{M}>0$. Since $M$ is the total bulk mass 
integrated up to the location of the brane,
the increase in $M$ implies an energy flow from the brane
to the bulk. Therefore, in this case, the energy in the
brane tension is transferred to the bulk scalar field and
it flows out into the bulk.
In contrast, if $\delta<(-\Delta)/8$, we have $\dot M<0$.
In this case, although there is still energy transfer from the
brane tension to the bulk scalar field, 
the bulk energy flows onto the brane.
In other words, there is a localization process of the
bulk energy onto the brane that overwhelms
the energy released from the brane tension.

\subsection{Local mass and the effective 4-dimensional description}

In the bulk inflaton model with a quadratic potential
\cite{Himemoto:2000nd, Minamitsuji:2003pb, Langlois:2003dd, Tanaka:2003eg},
it has been shown that
 the bulk scalar field projected on the brane behaves exactly
like a 4-dimensional field in the low energy limit, $H^2\ell^2\ll1$,
where $H$ is the Hubble parameter of the brane, and the leading
order correction gives the gradual energy loss from the scalar field
to the bulk, giving rise to the dark radiation 
term~\cite{Himemoto:2000nd, Langlois:2003dd}.
Here, we discuss the relation between the dark radiation term
appearing in this effective 4-dimensional description and
the generalized dark radiation term given by the local mass
in the bulk.

{}From the geometrical description~\cite{Shiromizu:1999wj}, the induced
Einstein equation on the brane is written as 
\begin{eqnarray}
{}^{(4)}G_{\mu\nu}= -{1 \over 12}\kappa_{5}^4 \sigma^2 q_{\mu\nu}
+\kappa_{5}^2\tilde{T}^{(b)}_{\mu\nu}-E_{\mu\nu},
\end{eqnarray}
where 
\begin{eqnarray}
\tilde{T}^{(b)}_{\mu\nu}=
{2\over 3}\Bigl[\tilde{T}_{ab}q^a_{\mu}q^{b}_{\nu}
-\Bigl(\tilde{T}_{ab}n^a n^b
-{1\over 4}\tilde{T}_{ab}g^{ab}\Bigr)q_{\mu\nu}\Bigr],
\end{eqnarray}
is the projected tensor of the bulk energy-momentum onto the brane
which includes the contribution of the cosmological constant; see
Eq.~(\ref{tildeT}).
For a homogeneous and isotropic brane, the non-vanishing components are
\begin{eqnarray}
&&\tilde{T}^{(b)}_{\tau\tau}=
-\tilde{T}^v{}_{v}
+{1\over 6}\tilde{T}^{i}{}_{i},
\nonumber\\
&&\tilde{T}^{(b)\,i}{}_{i}=
{1\over 6}\tilde{T}^i{}_{i}
-{\dot{u}\over \dot{v}}\tilde{T}^v{}_{u}
-{\dot{v}\over \dot{u}}\tilde{T}^u{}_{v}
+\tilde{T}^{v}{}_{v}.
\end{eqnarray}

Let us decompose $E_{\mu\nu}$ as
\begin{eqnarray}
E_{\mu\nu}=E^{(b)}_{\mu\nu}+E^{(d)}_{\mu\nu},
\end{eqnarray}
where $E^{(b)}_{\mu\nu}$ is to be expressed in terms of the
bulk energy-momentum tensor in such a way that the effective
4-dimensional description is recovered, and $E^{(d)}_{\mu\nu}$
is the part that should be identified as the dark radiation term
in this effective 4-dimensional approach.
To be in accordance with~\cite{Langlois:2003dd}, 
we choose the components of
$E^{(b)}_{\mu\nu}$ as
\begin{eqnarray}
-E^{(b)}_{tt}
=-{1\over 8}\kappa_{5}^2
\Bigl({\dot{u}\over \dot{v}}\tilde{T}^v{}_{u}
+{\dot{v}\over\dot{u}}\tilde{T}^u{}_{v}\Bigr) 
+{1\over 12}\kappa_{5}^2\Bigl(\tilde{T}^{i}{}_{i}
-3\tilde{T}^{v}{}_{v}\Bigr)
=-E^{(b)\,i}{}_{i},
\end{eqnarray}
and identify the remaining part with the dark radiation
term, $X$,
\begin{eqnarray}
-E^{(d)}_{tt}=X=-E^{(d)\,i}{}_{i}.
\end{eqnarray}

In the effective 4-dimensional description, 
the Einstein equation on the brane takes the form,
\begin{eqnarray}
{}^{(4)}G_{\mu\nu}
=\kappa_{4}^2T^{\rm (eff)}_{\mu\nu}-E^{(d)}_{\mu\nu},
\end{eqnarray}
where $T^{\rm (eff)}_{\mu\nu}$ is the effective energy-momentum
tensor on the brane,
\begin{eqnarray}
\kappa_{4}^2T^{\rm (eff)}_{\mu\nu}= 
-{1 \over 12}\kappa_{5}^4 \sigma^2 q_{\mu\nu}
+\kappa_{5}^2\tilde{T}^{(b)}_{\mu\nu}-E^{(b)}_{\mu\nu}\,,
\end{eqnarray}
and $\kappa_4^2$ is the 4-dimensional gravitational constant
that should be appropriately defined to agree with
the conventional 4-dimensional Einstein gravity in the low
energy limit.
In the present case of homogeneous and isotropic
cosmology, the only non-trivial components are
the effective energy density and pressure, which are
 given explicitly by
\begin{eqnarray}
&&\kappa_{4}^2\rho^{({\rm eff})}
=-\kappa_{4}^2\tilde{T}^{({\rm eff}) \tau}{}_{\tau}
={1\over 12 }\kappa_{5}^4\sigma^2
-{5\over 4}\kappa_{5}^2\tilde{T}^{v}{}_{v}
+{1\over 4}\kappa_{5}^2\tilde{T}^{i}{}_{i}
-{1\over 8}\kappa_{5}^2\Bigl({\dot{u}\over \dot{v}}
\tilde{T}^{v}{}_{u}
+{\dot{v}\over \dot{u}}\tilde{T}^{u}{}_{v}\Bigr),
 \nonumber \\
&& \kappa_{4}^2p^{({\rm eff})}
={1\over 3}\kappa_{4}^2\tilde{T}^{({\rm eff})\,i}{}_{i}
= -{1\over 12}\kappa_{5}^4\sigma^2 
+{1\over 4}\kappa_{5}^2 \tilde{T}^{v}{}_{v}
+{1\over 12}\kappa_{5}^2\tilde{T}^{i}{}_{i}
-{3\over 8}\kappa_{5}^2 \Bigl( {\dot{u}\over \dot{u}}
 \tilde{T}^{v}{}_{u}
+{\dot{v}\over \dot{u}} \tilde{T}^{u}{}_{v}\Bigr).
\label{effrhop}
\end{eqnarray}
The effective Friedmann equation on the brane is written as
\begin{eqnarray}
3\Bigl[H^2+{K\over r^2}\Bigr]
&=&\kappa_{4}^2 \rho^{({\rm eff})}-E^{(d)}_{\tau\tau}
\nonumber \\
&=&{1\over 12 }\kappa_{5}^4\sigma^2
-{5\over 4}\kappa_{5}^2\tilde{T}^{v}{}_{v} 
+{1\over 4}\kappa_{5}^2\tilde{T}^{i}{}_{i}
-{1\over 8}\kappa_{5}^2 \Bigl(
{\dot{u}\over \dot{v}}\tilde{T}^{v}{}_{u}
+{\dot{v}\over \dot{u}}\tilde{T}^{u}{}_{v}
 \Bigr) +X.
\end{eqnarray}
Applying this to a bulk scalar field with the action~(\ref{bulkscalar}),
 we find $\rho^{(\rm eff)}$ and $p^{(\rm eff)}$
are given by those of a 4-dimensional scalar field $\varphi$
 with the potential,
\begin{eqnarray}
\kappa_4^2V^{(\rm eff)}(\varphi)=
\kappa_5^2\left(\frac{1}{2}V(\phi)
+{\kappa_5^2\over12}\sigma^2(\phi)-{1\over16}\sigma'{}^2(\phi)\right),
\end{eqnarray}
where $\varphi=\sqrt{\kappa_5^2/\kappa_4^2}\,\phi$.
{}From the contracted Bianchi identity, we obtain
\begin{eqnarray}
D^{\mu}T^{({\rm eff})}_{\mu\tau}
=\Bigl[\dot\rho^{({\rm eff})}
+3H \Bigl(\rho^{({\rm eff})}+p^{({\rm eff})}\Bigr) \Bigr]
=-\frac{1}{\kappa_5^2} \frac{1}{r^4}(r^4 X)\dot{}\,.
\end{eqnarray}
Unfortunately, as we can see from Eqs.~(\ref{effrhop}),
there is no simplification in the energy equation in terms of 
the 5-dimensional energy-momentum tensor.

{}From the effective 4-dimensional point of view,
what happens is the conversion of the scalar field energy
on the brane to the dark radiation via the
coupling to the brane tension.
{}From the bulk point of view, a natural interpretation
is to regard the local mass $M$ on the brane
as the generalized dark radiation. 
These two different identifications of the dark radiation term on
the brane coincide only when the bulk is in vacuum and
$M$ is constant. Comparison of the above decomposition of
$E_{\tau\tau}$ with Eq.~(\ref{Etautau}), we
find the difference between the dark radiation in the 4-dimensional
description and the generalized dark radiation in terms of the local
mass $M$ as
\begin{eqnarray}
{3\tilde{M}\over r^4}=
{1\over 4}\kappa_{5}^2\Bigl(
\tilde{T}^{i}_{\,\,\,i}-5\tilde{T}^{v}_{\,\,\,v}\Bigr)
-{1\over 8}\kappa_{5}^2\Bigl({\dot{u}\over \dot{v}}\tilde{T}^{v}_{\,\,\,u}
+{\dot{v}\over \dot{u}}\tilde{T}^{u}_{\,\,\,v}\Bigr)
+X.
\end{eqnarray}

\end{document}